\documentclass[12pt,draftcls,onecolumn]{IEEEtran}

\usepackage{amsthm,amsmath,amssymb}
\usepackage{eucal}
\usepackage{xifthen}
\usepackage{mathtools}
\usepackage{enumerate}
\usepackage{microtype}
\usepackage{xspace}
\usepackage{bm}
\usepackage{cite}
\usepackage{fancyhdr}
\usepackage{lastpage}
\usepackage[small]{caption}
\usepackage{xcolor}
\usepackage{algorithm}
\usepackage{algpseudocode}
\allowdisplaybreaks

\long\def\comment#1{}

\newfont{\bbb}{msbm10 scaled 700}

\newfont{\bb}{msbm10 scaled 1100}
\newcommand{\CC}{\mbox{\bb C}}

\newcommand{\RR}{\mbox{\bb R}}

\newcommand{\mbs}[1]{\bm{#1}}
\newcommand{\vect}[1]{{\lowercase{\mbs{#1}}}}
\newcommand{\mat}[1]{{\uppercase{\mbs{#1}}}}

\newcommand{\Pmatrix}[1]{\begin{array}{ll}#1\end{array}}

\newcommand{\T}{{\scriptscriptstyle\mathsf{T}}}
\renewcommand{\H}{{\scriptscriptstyle\mathsf{H}}}

\renewcommand{\Re}[1][]{\ifthenelse{\isempty{#1}}{\operatorname{Re}}{\operatorname{Re}\left(#1\right)}}
\renewcommand{\Im}[1][]{\ifthenelse{\isempty{#1}}{\operatorname{Im}}{\operatorname{Im}\left(#1\right)}}

\newcommand{\cv}{\vect{c}}

\newcommand{\ev}{\vect{e}}

\newcommand{\gv}{\vect{g}}

\newcommand{\nv}{\vect{n}}

\newcommand{\rv}{\vect{r}}
\newcommand{\sv}{\vect{s}}

\newcommand{\vv}{\vect{v}}
\newcommand{\wv}{\vect{w}}
\newcommand{\xv}{\vect{x}}

\newcommand{\zerov}{\vect{0}}

\newcommand{\psiv}{\hbox{\boldmath$\psi$}}

\newcommand{\Psim}{\hbox{\boldmath$\Psi$}}

\newcommand{\Am}{\mat{a}}
\newcommand{\Bm}{\mat{b}}
\newcommand{\Cm}{\mat{c}}

\newcommand{\Pm}{\mat{p}}
\newcommand{\Qm}{\mat{q}}

\newcommand{\Tm}{\mat{t}}

\newcommand{\Xm}{\mat{x}}
\newcommand{\Ym}{\mat{y}}

\newcommand{\Ac}{{\mathcal A}}

\newcommand{\Cc}{{\mathcal C}}

\newcommand{\Ec}{{\mathcal E}}

\newcommand{\Gc}{{\mathcal G}}

\newcommand{\Ic}{{\mathcal I}}
\newcommand{\Jc}{{\mathcal J}}

\newcommand{\Nc}{{\mathcal N}}

\newcommand{\Pc}{{\mathcal P}}

\newcommand{\Rc}{{\mathcal R}}
\newcommand{\Sc}{{\mathcal S}}
\newcommand{\Tc}{{\mathcal T}}

\newcommand{\Vc}{{\mathcal V}}

\newcommand{\Id}{\mat{\mathrm{I}}}
\newcommand{\one}{\mat{\mathrm{1}}}

\newcommand{\CN}[1][]{\ifthenelse{\isempty{#1}}{\mathcal{N}_{\mathbb{C}}}{\mathcal{N}_{\mathbb{C}}\left(#1\right)}}
\renewcommand{\P}[1][]{\ifthenelse{\isempty{#1}}{\mathbb{P}}{\mathbb{P}\left(#1\right)}}
\newcommand{\E}[1][]{\ifthenelse{\isempty{#1}}{\mathbb{E}}{\mathbb{E}\left(#1\right)}}
\renewcommand{\det}[1][]{\ifthenelse{\isempty{#1}}{\mathrm{det}}{\mathrm{det}\left(#1\right)}}
\newcommand{\trace}[1][]{\ifthenelse{\isempty{#1}}{\mathrm{tr}}{\mathrm{tr}\left(#1\right)}}
\newcommand{\rank}[1][]{\ifthenelse{\isempty{#1}}{\mathrm{rank}}{\mathrm{rank}\left(#1\right)}}
\newcommand{\diag}[1][]{\ifthenelse{\isempty{#1}}{\mathrm{diag}}{\mathrm{diag}\left(#1\right)}}

\DeclarePairedDelimiter\abs{\lvert}{\rvert}
\DeclarePairedDelimiter\Abs{\lvert}{\rvert^2}
\DeclarePairedDelimiter\norm{\lVert}{\rVert}
\DeclarePairedDelimiter\Norm{\lVert}{\rVert^2}

\renewcommand{\arg}{{\hbox{arg}}}

\renewcommand{\Re}{{\rm Re}}
\renewcommand{\Im}{{\rm Im}}

\allowdisplaybreaks
\newcommand{\st}{{\rm s.t.}}
\DeclareMathAlphabet{\mathcal}{OMS}{cmsy}{m}{n}

\newcommand{\defeq}{\triangleq}

\newtheorem{definition}{Definition}
\newtheorem{example}{Example}

\usepackage{hyperref}
\hypersetup{
    bookmarks=true,         
    unicode=false,          
    pdftoolbar=true,        
    pdfmenubar=true,        
    pdffitwindow=false,     
    pdfstartview={FitH},    
    pdfnewwindow=true,      
    colorlinks=true,       
    linkcolor=red,          
    citecolor=green,        
    filecolor=magenta,      
    urlcolor=cyan           
}

\usepackage{algpseudocode}

\newcommand{\p}{{\rm p}}
\newcommand{\dd}{{\rm d}}

\begin{document}
\title{Topological Pilot Assignment in Large-Scale Distributed MIMO Networks}
\author{
\IEEEauthorblockN{Han Yu, Xinping Yi, and Giuseppe Caire}
\thanks{H.~Yu and X.~Yi are with Department of Electrical Engineering and Electronics at The University of Liverpool, Liverpool L69 3BX, United Kingdom (email: {\tt xinping.yi@liverpool.ac.uk})}
\thanks{G.~Caire is with Department of Electrical Engineering and Computer Science at Technical University of Berlin, 10587 Berlin, Germany (email: {\tt caire@tu-berlin.de})}
}

\maketitle

\begin{abstract}
We consider the pilot assignment problem in large-scale distributed multi-input multi-output (MIMO) networks, where a large number of remote radio head (RRH) antennas are randomly distributed in a wide area, and jointly serve a relatively smaller number of users (UE) coherently. By artificially imposing topological structures on the UE-RRH connectivity, we model the network by a partially-connected interference network, so that the pilot assignment problem can be cast as a topological interference management problem with multiple groupcast messages. Building upon such connection, we formulate the topological pilot assignment (TPA) problem in two different ways with respect to whether or not the to-be-estimated channel connectivity pattern is known {\em a priori}. When it is known, we formulate the TPA problem as a low-rank matrix completion problem that can be solved by a simple alternating projection algorithm. Otherwise, we formulate it as a sequential maximum weight induced matching problem that can be solved by either a mixed integer linear program or a simple yet efficient greedy algorithm. With respect to two different formulations of the TPA problem, we evaluate the efficiency of the proposed algorithms under the cell-free massive MIMO setting.
\end{abstract}


\section{Introduction}  \label{sec:intro}

The last decades have witnessed the advances of multiple-user multiple-input multiple-output (MIMO) technologies towards the next generation wireless communications systems (e.g., 5G and beyond), particularly in terms of antenna array from small size to massive MIMO, in terms of duplex operations from frequency-division duplex (FDD) to time-division duplex (TDD), and in terms of network architectures from centralized (e.g., cloud) to distributed (e.g., fog) radio access networks. 
As one of the key wireless access techniques in 5G and beyond, massive MIMO promises high-throughput and low-latency services with low-complexity transceivers.

Conventional massive MIMO makes use of a collocated antenna array with a large number (e.g., hundreds) of elements at the base station (BS), which coherently serve a relatively smaller number (e.g., tens) of users (UEs) in the same time-frequency resource \cite{mMIMO,mMIMO-SPM}. In doing so, users' channels exhibit some interesting properties: channel hardening and favorable propagation. Because of the large antenna array, the average channel gains across time between the BS and the UEs are almost deterministic, ruling out the small-scale fading effects that are disadvantageous to high-throughput wireless services. 
The large antenna array also provokes the favorable propagation channels in the sense that the users with distinct angle-of-arrival have almost orthogonal channel vectors. This lends itself to the use of conjugate beamforming instead of the more sophisticated zero-forcing beamformers, and thus the transceivers design can be significantly simplified. 

Recently, a distributed deployment spreading out the massive number of antennas over a large area demonstrates the superior network performance over the collocated counterpart. The motivation of such distributed massive MIMO is two-fold. On one hand, distributed wireless access networks promises potentially higher coverage thanks to the exploitation of the diversity against shadow fading. On the other hand, the emerging applications such as Internet of Things encourage smart devices with distributed locations to be potential RRHs, so that a distributed antenna deployment sounds more promising for ubiquitous communications in the future. 

Most recently, various distributed network architectures for massive MIMO have been proposed with different focuses. For instance, cell-free massive MIMO \cite{Cell-Free,Cell-Free-ComM} promotes the ``cell-free'' concept in which every UE will be jointly served by all RRHs so that no handover will incur when the UE moves because it is always within the single huge cell. A central processing unit is enabled to coordinate information exchange and jointly compute system parameters (e.g., channel estimation and power control). Such a ``cell-free'' concept has attracted a lot of attention recently, including the considerations of spectral and energy efficiency \cite{ngo2017total,chen2018channel}, precoding and power optimization \cite{CF-power,bashar2019uplink}, limited-capacity fronthaul \cite{bashar2019max}, user-centric approaches \cite{buzzi2017cell}, the mmWave scenario \cite{alonzo2019energy}, among many others (see a comprehensive survey \cite{chen2021survey} and references therein).
On the other hand, the ``fog'' massive MIMO proposed in \cite{Fog-mMIMO} is dedicated to a seamless and implicit user association architecture in which the UEs are assigned to certain RRHs with large-scale antenna array in an autonomous manner by a novel coded ``on-the-fly'' pilot contamination control mechanism, leading to autonomous handover as UE moves and thus establishing a cell-transparent network.
{A problem with both cell-free and fog architectures is that, since there is no clear cell boundary any longer, the uplink pilot assignment to the UEs is done once for all and not re-assigned at every handover as simplicity assumed in cellular-based massive MIMO systems in order to guarantee that intra-cell users have mutually orthogonal uplink pilots \cite{mMIMO}. 
Hence, the inherent pilot contamination due to non-orthogonal pilots represents an important limiting factor that is handled by global pilot optimization in the cell-free scheme \cite{Cell-Free} or with coding and ``on-the-fly'' contamination control in the fog scheme \cite{Fog-mMIMO}.

To address the pilot contamination issue, a number of works have concentrated on low-complexity pilot assignment algorithms in the cell-free massive MIMO setting. 
In particular, a greedy pilot assignment method was proposed in \cite{Cell-Free} to gradually refine the random assignment in order to gain improved throughput performance. A dynamic pilot reuse scheme was proposed in \cite{sabbagh2018pilot} by using user-centric clustering methods. By modeling the conflict of pilot assignment between UEs as an interference graph, graph coloring based methods (e.g., \cite{liu2020graph,hmida2020graph,masoumi2020cell}) were proposed for pilot assignment.  
The joint RRH selection and pilot assignment was considered in \cite{bjornson2020scalable} to make the network more scalable, and structured policies were proposed in \cite{attarifar2018random,chen2020structured} together with clustering techniques (e.g., K-means and user grouping).
The pilot assignment can be also formulated as a graph matching problem \cite{buzzi2021pilot}, which can be solved efficiently by Hungarian algorithm. A heap-based algorithm has been adopted in \cite{nguyen2020spectral} to reduce pilot contamination and enhance spectral efficiency, and a tabu search method in \cite{liu2019tabu} to exploit local neighborhood search.  
Although promising, these approaches either rely on sum rate evaluation during the pilot assignment process, or on heuristics without theoretical guarantees.
In the former, rate calculation also involves power allocation and channel estimation, which is related to pilot assignment. This is a ``chicken-and-egg'' problem. In the latter, although some heuristics work well in small-scale networks, they are not provably scalable for large-scale ones.
As the pilot assignment problem has a combinatorial nature, it is in general NP-hard and challenging to find a provably scalable solution. In this regard, a natural question then arises as to how we can come up with a principled way for pilot assignment by making use of only the long-term channel information.

As a matter of fact, inspecting such distributed massive MIMO networks, one may notice that some previously ignored UE-RRH connectivity patterns may be exploitable and of great benefit. Owing to the random locations of RRH antennas, the fact that power decays rapidly with distance, the existence of obstacles, and local shadowing effects, we may argue that certain UE-RRH links are unavoidably much weaker than others,  which by intuition makes these concerned RRHs not suitable to serve some UEs. This is also confirmed by the simulations in e.g., \cite{Cell-Free,CF-power}, where only a small fraction of RRHs contribute most to a UE while the contribution of the rest is negligible. Thus, the channels with negligible contributions are not necessarily estimated, and one pilot sequence can be allocated to more UEs as long as it does not cause severe pilot contamination.
As such, it suggests the use of a partially-connected bipartite graph to model, at least approximately, the network connectivity, i.e., which RRH antenna serves which UE, to artificially sparsify the network topology and channel estimation pattern, so that the pilot assignment can be done based on the sparsified UE-RRH connectivity. 

In this paper, we focus on the pilot assignment problem in the distributed (e.g., cell-free or fog) massive MIMO systems, and aim to provide another perspective to investigate such a challenging problem. We impose a topological structure on the network connectivity based on the large-scale fading coefficients, so that only channels with larger path-loss than a certain threshold are captured and the network connectivity is artificially sparsified. Based on such a sparsified network topology, we connect the pilot assignment problem to the topological interference management (TIM) problem with multiple groupcast message setting, so that the developed coding schemes for TIM using e.g., graph coloring and coded multicasting, can be applied here for pilot assignment. Instead of analyzing the optimality with respect to specific topologies in TIM, we propose two systematic pilot assignment methods to deal with arbitrary topologies by formulating two non-convex optimization problems. The first one is a low-rank matrix completion formulation to minimize the pilot dimension with a given channel estimation pattern. In particular, it minimizes the rank of a partially determined matrix whose entries are determined by the channel estimation pattern and a binary pilot assignment matrix. { Once the matrix is completed, we employ matrix factorization to obtain the binary pilot assignment matrix.} 
The second one is a formulation of binary quadratically constrained quadratic program to find the optimal channel estimation pattern with a given training budget (i.e., pilot dimension). Instead of solving the problem directly, we apply the sequential optimization method to solve it iteratively, and at each iteration we solve a combinatorial optimization problem to maximize the usage of each pilot dimension, in the hope to estimate as many channels as possible. 
{By such formulation, we propose a mixed integer program formulation via sequential maximum weight induced matching and a simple yet efficient greedy algorithm.}
The superiority of two proposed methods are verified by Monte-Carlo simulation under the cell-free massive MIMO settings, which show that our approaches have a better ergodic rate performance compared to the state-of-the-art methods.

\underline{\bf Notation}: 
Throughout this paper, we abbreviate $[n] \defeq \{1,2,\dots,n\}$ for an integer $n$. $[\Am]_{ij}$ presents the $ij$-th entry of the matrix $\Am$, and $\Am_{\Ic,\Jc}$ denotes the submatrix of $\Am$ where $\Ic$ and $\Jc$ indicate the indices of selected rows and columns respectively. $\abs{\Ac}$ is the cardinality of the set $\Ac$. We denote by ${\bf 1}_{M \times 1}$ the all one $M \times 1$ vector, by $\Id_M$ the $M \times M$ identity matrix, and by $\ev_m$ the $m$-th column of the identity matrix.
{We abbreviate $\{a_t\}_{t} \defeq \{a_t, \; \forall t\}$ and for the multiple indices, it applies similarly.}

\section{System Modeling}
\label{sec:sys}
\subsection{Distributed Massive MIMO}
Consider a distributed massive MIMO network with $M$ remote radio heads (RRHs) each with single antenna\footnote{For ease of presentation, we focus on the single-antenna RRHs for the derivation, whereas the extension to multiple-antenna RRHs is straightforward.} coherently and simultaneously serving $K$ single-antenna user equipment (UEs), all of which are uniformly located in a large area at random. The RRHs operate in TDD mode, so that the downlink channel coefficients can be estimated through uplink training due to the uplink/downlink channel reciprocity in TDD mode. All RRHs are connected to a central processing unit (CPU) via error-free backhaul links for the purpose of coordination. The backhaul links are not allowed to exchange instantaneous channel state information (CSI), while payload data, pilot assignment strategy, and power control coefficients can be routed and exchanged. It is assumed $M\gg K$, and each UE should be served by a sufficiently large number of RRHs in order to harvest the benefits of channel hardening and favorable propagation. Through the limited coordination among RRHs, a distributed massive MIMO is formed. 

The channel coefficient $g_{mk}$ between RRH-$m$ and UE-$k$ is modeled as follows:
\begin{align} \label{channel-coefficient}
g_{mk} = \sqrt{\beta_{mk}} h_{mk},
\end{align}
where $\beta_{mk}$ is the large-scale fading (i.e., path-loss) coefficient, and $h_{mk}$ is small-scale fading and is assumed to be a complex i.i.d. Gaussian random variable with mean 0 and variance 1 (i.e., $\Cc\Nc(0,1)$).
The channel coefficients are assumed to be constant during a TDD frame. A TDD frame consists of UL training and DL payload transmission. In this work, we place our focus mainly on pilot assignment and channel estimation.

\subsection{Uplink Training}

Let $\tau_\p$ be the {{\em maximal} duration (in samples) reserved for UL training phase, during which each UE is assigned with a combination of orthogonal pilot sequences $\{\psiv_t \in \CC^{T \times 1}, t \in [T]\}$ with $T \le \tau_p$ being the pilot dimension {\em actually} used for UL training. We impose $\psiv_t^\H \psiv_s =\delta(t,s)$ to ensure the pilot orthogonality.} Note that $T$ can be much less than the number of users $T<K$, where a pilot sequence can be reused by multiple users with proper pilot contamination control.

For a specific $\psiv_t$, we introduce a set of binary variables
\begin{align}
x_{kt} = \left\{\Pmatrix{1, & \text{if UE-$k$ is assigned $\psiv_t$ with success}\\0, & \text{otherwise.}}\right.,
\end{align}
{so that the pilot signal sent from UE-$k$ can be specified by
\begin{align}
\sv_{\p,k} = \sqrt{\tau_\p \eta_{\p}}\sum_{t=1}^T x_{kt} \psiv_t,
\end{align}
where $\eta_{\p}$ is the normalized power coefficient such that 
\begin{align} \label{eq:pilot-power-constraint}
\frac{1}{K}\sum_{k=1}^K\E_{}[\Norm{\sv_{\p,k}}] \le \tau_\p \rho_{\p}
\end{align}
with $\tau_\p \rho_{\p}$ being the average power reserved for each UE over UL training.} For equal pilot power allocation, we have $\eta_{\p}=\frac{K\rho_{\p}}{\sum_{k=1}^K\sum_{t=1}^T x_{kt}}$.

At the $m$-th RRH, the received pilot signal over $T$ pilot dimensions can be given by
\begin{align}
\rv_{\p,m} &=  \sum_{k=1}^K g_{mk} \sv_{\p,k} + \wv_{\p,m}\\
&= \sqrt{\tau_\p \eta_{\p}} \sum_{k=1}^K \sum_{t=1}^T g_{mk} x_{kt} \psiv_t + \wv_{\p,m}
\end{align}
where $\wv_{\p,m} \in \CC^{T \times 1}$ is the additive white Gaussian noise (AWGN) at RRH-$m$, and is i.i.d. over $T$ with $\Cc\Nc(0,\Id_T)$.

Given the above pilot signal, the RRHs check every pilot dimension and try to estimate  certain channels.
At the $m$-th RRH, the received pilot signal is multiplied by every pilot sequence $\psiv_{t}$ to estimate the channels from some UE-$k$ to RRH-$m$. Thus, the resulting pilot signal {observed at the output of the $t$-th pilot correlator $\hat{r}_{\p,mt} = \psiv_{t}^\H  \rv_{\p,m}$ can be written as}
\begin{align} \label{estimated-signal}
\hat{r}_{\p,mt} 
&=  \sqrt{\tau_\p \eta_{\p}} \sum_{k=1}^K g_{mk} x_{kt} +  \psiv_{t}^\H \wv_{\p,m}  \\
&=  \sqrt{\tau_\p \eta_{\p}}  g_{mk} x_{kt} + \sqrt{\tau_\p \eta_{\p}}  \sum_{k' \neq k} g_{mk'}x_{k't} +  \psiv_{t}^\H \wv_{\p,m} 
\nonumber
\end{align}
{The next step consists of recovering} $\{g_{mk}\}_{m,k}$ from the received pilot signals and obtain the corresponding estimates $\{\hat{g}_{mk}\}_{m,k}$. A channel estimate is stable in the sense that mean square error (MSE) satisfies $\E {[\Abs{g_{mk}-\hat{g}_{mk}}]} \to 0$ when $\rho_{\p} \to \infty$.
The channel coefficient $g_{mk}$ can be estimated using different estimators, such as least square (LS), minimum mean square error (MMSE).
For instance, the MMSE estimate of $g_{mk}$ can be produced by 
\begin{align} \label{eq:channel-estimate}
\hat{g}_{mk} &= \frac{\E \big[ \hat{r}_{\p,mt}^\H {g}_{mk} \big] }{\E \big[\Abs{\hat{r}_{\p,mt}}\big]} \hat{r}_{\p,mt} 
= \frac{\sqrt{\tau_\p \eta_{\p}} \beta_{mk} x_{kt} }{1+\tau_\p \eta_{\p}  \sum_{k'} \beta_{mk'} x_{k't}}  \hat{r}_{\p,mt}
\end{align}
for some $t$.
The MSE, for which RRH-$m$ estimates the channel coefficient $g_{mk}$ through pilot $\psiv_t$ when UE-$k$ is sending pilot $\psiv_t$ as well, can be written as
\begin{align}
\text{MSE}_{mkt} &= \E \{ \Abs{{g}_{mk}}\} - \frac{\Abs{\E \{ \hat{r}_{\p,mt}^\H {g}_{mk} \}}}{\E \{\Abs{\hat{r}_{\p,mt}}\}} \\
&= \beta_{mk} - \frac{\tau_\p \eta_{\p} \beta_{mk}^2 x_{kt} }{1 + \tau_\p \eta_{\p}  \sum_{k'} \beta_{mk'} x_{k't} }. \label{eq:mse-general}
\end{align}
Apparently, obtaining a meaningful estimate of $g_{mk}$ requires $x_{kt}=1$ and $x_{k't}=0$ for all $k' \neq k$. That is, UE-$k$ is assigned pilot $\psiv_{t}$ exclusively, so that $g_{mk}$ can be stably estimated at RRH-$m$ by using $\psiv_{t}$ with diminishing estimation error as $\rho_{\p}$ tends to infinity. If the UE-RRH connectivity is equally strong for any pair of UE and RRH, the stable estimate of all channels requires that each UE is assigned a unique orthogonal pilot sequence, so that the total pilot dimension is at least $K$.

Nevertheless, we argue that under the distributed MIMO setting, it is unnecessary to estimate all channel coefficients between every RRH and every UE; rather, the UE-RRH links with negligible contributions can be ignored.
As such, over $T$ pilot dimension, let ${\Tc}_{E,m}$ represent the indices of UEs whose channels are stably estimated at RRH-$m$, and ${\Rc}_{E,k}$ represent the indices of RRHs that are supposed to serve UE-$k$. 
While ${\Tc}_{E,m}$ is a consequence of pilot assignment, ${\Rc}_{E,k}$ is a system choice that determines the distribution of UEs' data across RRHs. In general, they are not necessarily related.

We hereafter refer to the channel estimation pattern specified by $\{{\Tc}_{E,m}\}_m$ as a bipartite graph $\Gc_E=([K],[M],\Ec_E)$ with the edge set 
\begin{align}
\Ec_E = \{(k,m): k \in {\Tc}_{E,m}, \forall m\}. 
\end{align}
As a first attempt, in this work we assume those RRHs who are supposed to serve UE-$k$ should possess stable estimates of the corresponding channel coefficients associated to UE-$k$, and those UEs whose channels are stably estimated by RRH-$m$ should be served by RRH-$m$.  
That is, $m \in {\Rc}_{E,k}$ if and only if $k \in {\Tc}_{E,m}$.
It follows that the edge set of $\Gc_E$ can be alternatively represented by the UE association pattern $\Ec_E= \{(k,m): m \in {\Rc}_{E,k}, \forall k\}$.

\subsection{Downlink Data Transmission}
Given the channel estimates $\{\hat{g}_{mk}\}_{k \in {\Tc}_{E,m}}$ at RRH-$m$, conjugate beamforming is employed to transmit the symbols $\{q_k\}_{k \in {\Tc}_{E,m}}$ to the UE-$k$. The transmitted signal from RRH-$m$ can be written by
\begin{align}
s_{\dd,m} = \sqrt{\rho_\dd} \sum_{k \in {\Tc}_{E,m}} \eta_{mk}^{1/2} \hat{g}_{mk}^* q_k
\end{align}
where $q_k$ is the desired symbol by UE-$k$ satisfying $\E{[\Abs{q_k}]}=1$, and $\eta_{mk}$ is the power allocation efficient associated to the transmitted symbol $q_k$ from RRH-$m$, subject to the average power constraint at each RRH $$\frac{1}{M}\sum_{m=1}^M\E {[\Abs{s_{\dd,m}}]} \le \rho_\dd.$$ According to the transmitted signal, the power constraint can be rewritten as
\begin{align}
\frac{1}{M}\sum_{m=1}^M\sum_{k \in {\Tc}_{E,m}} \eta_{mk} \gamma_{mk} \le 1
\end{align}
where $\gamma_{mk} \defeq \E{[\Abs{ \hat{g}_{mk}}]}$. Thus, the received signal at UE-$k$ is given by
\begin{align}
r_{\dd,k} &= \sum_{m=1}^M g_{mk} s_{\dd,m} + w_{\dd,k}  \\
&= \sqrt{\rho_\dd} \sum_{m \in {\Rc}_{E,k}} \eta_{mk}^{1/2} g_{mk} \hat{g}_{mk}^* q_k + \sqrt{\rho_\dd} \sum_{m=1}^M \sum_{k' \neq k, k' \in {\Tc}_{E,m}} \eta_{mk'}^{1/2} g_{mk} \hat{g}_{mk'}^* q_{k'} + w_{\dd,k} \nonumber\\
&= f_{k,k} q_k + \sum_{k': k' \neq k}^K f_{k,k'} q_{k'} + w_{\dd,k} 
\end{align} 
where
\begin{align} \label{eq:equi-channel-coeff}
f_{k,k'} \defeq \sqrt{\rho_\dd} \sum_{m \in {\Rc}_{E,k'}} \eta_{mk'}^{1/2} g_{mk} \hat{g}_{mk'}^*.
\end{align}
Thus, the downlink received signal can be seen as an interference channel with channel coefficients $\{f_{k,k'}\}_{k,k'}$.
For simplicity, we assume that all channel coefficients in \eqref{eq:equi-channel-coeff} are known to the receivers. Taking into account the uplink training overhead, we have the {downlink} ergodic rate \cite{caire2018ergodic}
\begin{align}
R_k &= (1-\frac{T}{N_c})\E \left[ \log \left(1+\frac{\Abs{f_{k,k}}}{N_0+\sum_{k' \neq k} \Abs{f_{k,k'}}}\right) \right]
\end{align}
where $N_c$ is length of the TDD frame in samples, and $N_0$ is the normalized noise power.

\section{Topological Pilot Assignment}
\subsection{Topological Modeling}
Due to the fact that signal power decays fast as the distance increases and the shadowing effects, some UE-RRH links are unavoidably weak than others and thus both their contributions to joint transmission or influence as interference are negligible. It suggests the use of a UE-RRH connectivity pattern to model this at least approximately.
Thus, we introduce another weighted bipartite graph $\Gc=([K],[M],\Ec)$ in Fig. \ref{fig:pilot-assign} (Left) to represent the UE-RRH connectivity (i.e., network topology), where $[K]$ is the index set of UEs, $[M]$ is the index set of RRHs, and $\Ec$ is the collection of the edges with weights $\{\beta_{mk}\}_{m,k}$. The UE-$k$ is said to be connected to RRH-$m$, i.e., $(k,m) \in \Ec$, if and only if $\beta_{mk}\ge \delta_\beta$,  where the threshold $\delta_\beta$ is a crucial designing parameter. 
Let us denote by $\Tc_{m} \defeq \{k: (k,m) \in \Ec\}$ the indices of UEs connected to RRH-$m$ and by $\Rc_{k} \defeq \{m: (k,m) \in \Ec\}$ the indices of RRHs connected to UE-$k$. 

The network topology $\Gc$ captures both channel estimation pattern $\Gc_E$ that specifies the to-be-estimated channel pattern with significant contributions, and the non-negligible interference pattern that has negligible contributions to joint transmission yet non-negligible influence as interference.
Given $\Gc$ and $\Gc_E$, one may consider to impose such structures on pilot assignment problem in distributed massive MIMO systems. Hence, we formulate a topological pilot assignment (TPA) problem, dedicated to pilot assignment with artificially imposed network structures.
Without loss of generality, we assume $\Ec_E \subseteq \Ec$ that only strong channels should be estimated.

\begin{figure}
 \begin{minipage}{0.48\linewidth}
\includegraphics[width=1\columnwidth]{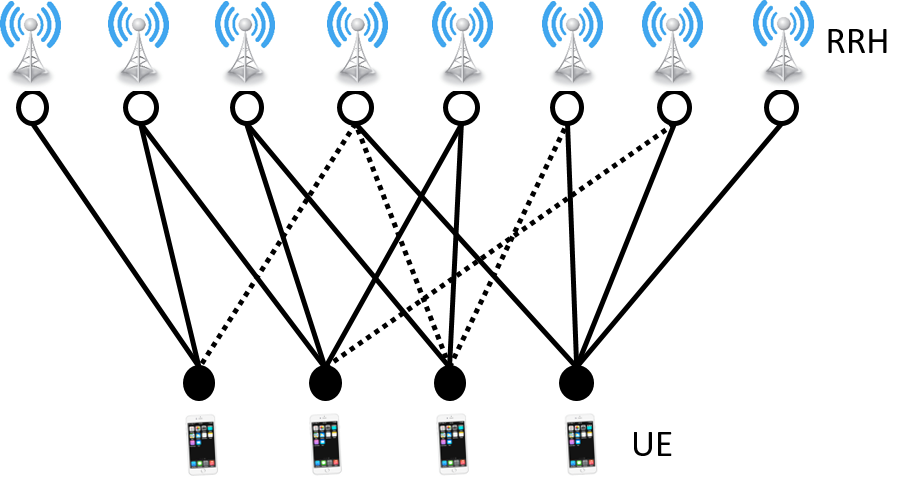}
\end{minipage}
 \begin{minipage}{0.48\linewidth}
\includegraphics[width=1\columnwidth]{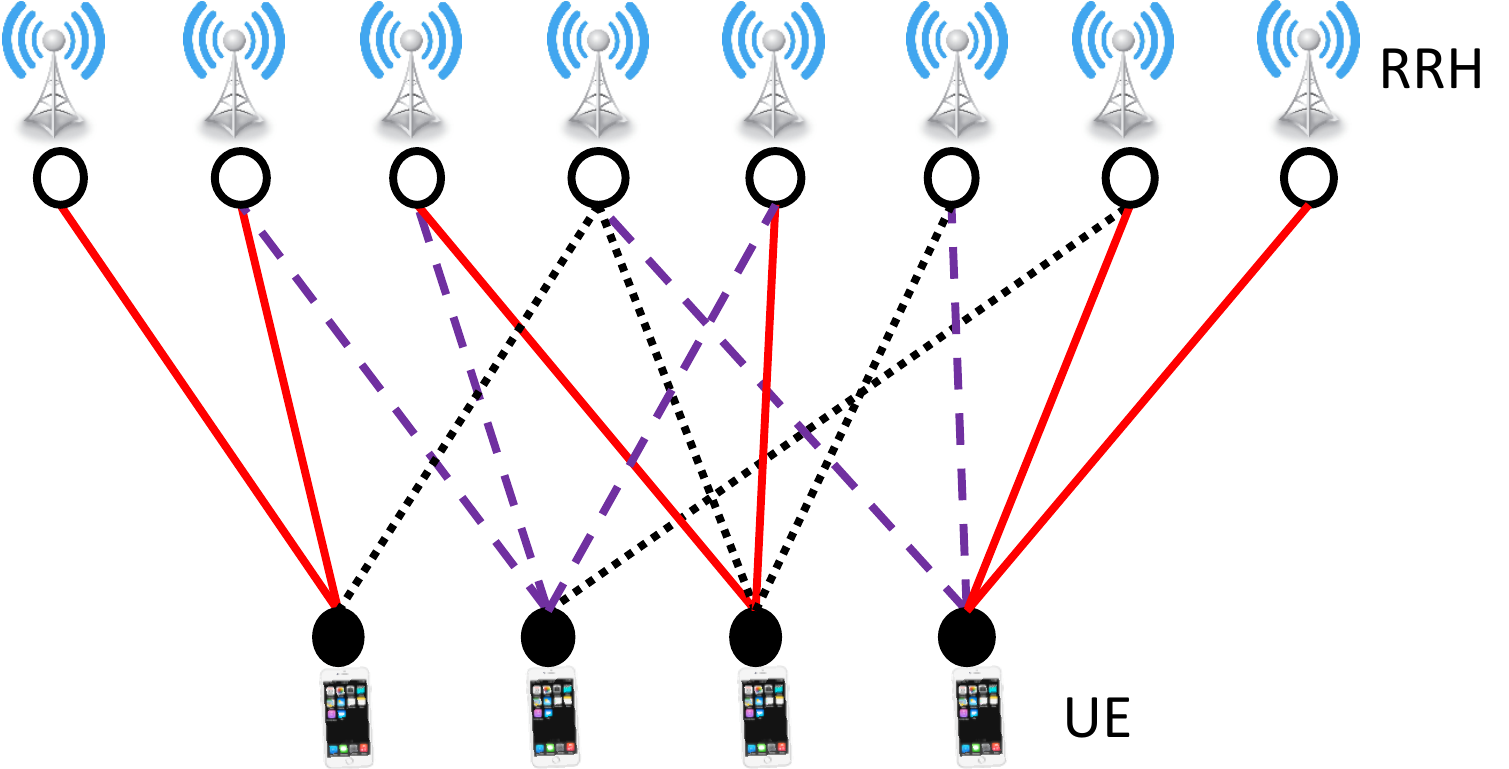}
\end{minipage}
\caption{Left: Topological modeling for a distributed massive MIMO network as a partially-connected bipartite graph, where all edges (including all solid and dotted ones) represent the UE-RRH connectivity, i.e., $\Ec(\Gc)$, and the solid edges represent the channel estimation pattern, i.e., $\Ec_E(\Gc_E)$. Right: A possible pilot assignment strategy, where different colors indicate distinct orthogonal pilot sequences. The colored edges cover the channel estimation pattern $\Ec_E(\Gc_E)$. By this pilot assignment, all users' channels of interest can be estimated stably because no pilot contamination is incurred at the RRHs.}
\label{fig:pilot-assign}
\vspace{-20pt}
\end{figure}

\begin{definition}
Given a UE-RRH connectivity pattern $\Gc=([K],[M],\Ec)$, the TPA problem consists of two subproblems:
\begin{itemize} 
\item {\em Pilot Dimension Minimization}, which focuses on allocating pilot sequences to minimize pilot dimension $T$ for a predetermined channel estimation pattern $\Gc_E$;
\item {\em Channel Pattern Optimization}, which is dedicated to determining the optimal channel estimation patterns $\Gc_E$ for a given pilot dimension $T$.
\end{itemize}
\end{definition}

It is worth noting that both subproblems rely highly on the choice of $\delta_\beta$ that determines the network topology $\Gc$. A larger $\delta_\beta$ makes the resulting topology sparser, so that a smaller $T$ is able to estimate all channels of the sparse network, while the uncaptured channels that are consequently not estimated may cause severe interference. On the contrary, a smaller $\delta_\beta$ leads to a denser network topology, so a specified pilot dimension may not able to estimate all channels of interest, while the non-estimated yet captured channels may cause severe degradation as well.

\subsection{Pilot Dimension Minimization}

The pilot dimension minimization subproblem aims to assign each UE a combination of orthogonal pilot sequences with minimal pilot dimension $T$ for a specified channel estimation pattern $\Gc_E$, so that all channels of interest can be properly and stably estimated. 
For instance, when UE-$k$ is using the pilot $\psiv_t$, any RRH-$m$ is supposed to be able to estimate the channel $g_{mk}$ if $(k,m) \in \Ec_E$ and the pilot signal at RRH-$m$ is not contaminated by other UEs using the same pilot $\psiv_t$. Meanwhile, for a specific RRH-$m$, any other UE-$j$ who has a strong channel connection to RRH-$m$, i.e., $(j,m) \in \Ec$ due to $\beta_{mj} \ge \delta_\beta$, is not supposed to use the same pilot $\psiv_t$ simultaneously. Otherwise the use of pilot $\psiv_t$ at both UE-$k$ and UE-$j$ will result in pilot contamination at RRH-$m$ so that the channels $g_{mk}$ cannot be stably estimated at RRH-$m$. 
\begin{example}
A feasible pilot assignment is shown in Fig. \ref{fig:pilot-assign} (Right), in which we assign two orthogonal pilots to estimate the channels of interest. In Fig. \ref{fig:pilot-assign} (Right), the edges in $\Ec_E$ are colored using two distinct colors, each of which represents an orthogonal pilot. Thus, given two orthogonal pilot sequences $\psiv_{1,2} \in \RR^{2 \times 1}$, UE-1 and UE-3 send $\psiv_1$, UE-2 sends the pilot $\psiv_2$, and UE-4 sends the combination of two pilots $\psiv_1+\psiv_2$. Then, RRH-$\{1,4,6,7,8\}$ see the uncontaminated pilot signal and can estimate the corresponding channels, whereas RRH-$\{2,3,5\}$ see the combination of two orthogonal pilot signals, and can estimate both channels stably over two timeslots by e.g., zero-forcing.
\end{example}

\subsection{Channel Pattern Optimization}
The channel pattern optimization subproblem is to decide which channel to be estimated given a total budget (e.g., pilot dimension) during the training phase.

Let us denote by $\Tc_{m} \defeq \{k: (k,m) \in \Ec\}$ the indices of UEs connected to RRH-$m$ and by $\Rc_{k} \defeq \{m: (k,m) \in \Ec\}$ the indices of RRHs connected to UE-$k$. Given two UE-$j,k$ such that $j,k \in \Tc_m$, the channels $g_{mk}$ and $g_{mj}$ cannot be estimated at RRH-$m$ using the same pilot sequence. That is, with a single pilot sequence, each RRH can only estimate at most one channel. On the other hand, given two RRH-$m,n$ such that $m,n \in \Rc_k$, the channels $g_{mk}$ and $g_{nk}$ can be estimated at RRH-$m$ and RRH-$n$ using the same pilot sequence. That is a single pilot could be used to estimate multiple channels originated from the same UE. As shown in Fig. \ref{fig:pilot-assign} (Right), for the pilot sequence denoted by red edges, each RRH estimates at most one channel and multiple channels may from the same UE.

The above rule yields the channel pattern that can be estimated by a single pilot sequence. Given a fixed pilot dimension (i.e., the number of orthogonal pilot sequences), the objective of this subproblem is to maximize the total number of channels to be estimated.

\subsection{Connection to Topological Interference Management}
A closer look at the TPA problem reveals the similarity to topological interference management (TIM) with message groupcasting \cite{Jafar:2013TIM}. Both TPA and TIM problems aim to exploit topological information for transmission in partially-connected interference networks without knowing channel coefficients at the transmitters. 

The TIM problem aims to deliver messages and the goal is to maximize the minimal  (symmetric) degrees of freedom $d_{\rm sym}$ achieved by all desired messages across all receivers. The groupcast message setting specifies that a message originated from a transmitter may be desired by multiple receivers, such that a message multicasting will benefit multiple receivers. In the TIM setting, $\Gc$ and $\Gc_E$ represent the network topology and desired message pattern respectively. 

The TPA problem aims to estimate the channel coefficients given the known pilot symbols, and the goal is to figure out how orthogonal pilot sequences are allocated to minimize the pilot dimensions $T$. It is feasible that all channels associated to one UE can be trained by one pilot sequence sent from this UE. In the TPA setting, $\Gc$ and $\Gc_E$ represent the network topology and channel estimation pattern respectively.

Intuitively, if we treat the channel coefficients in TPA as the symbols of the unknown messages in TIM, the pilot assignment in TPA can be obtained from the beamforming vectors of the encoding schemes for TIM with that additional constraint that they should be binary-valued.
Given a linear coding scheme for TIM groupcasting, we have translate it to a pilot assignment scheme for TPA, which yields $T=\frac{1}{d_{\rm sym}}$, where $d_{\rm sym}$ is the symmetric degrees of freedom under the TIM setting.
In light of such a connection, we can borrow the well-designed coding schemes from TIM to TPA. In what follows, we present two simple methods for the purpose of illustration: one is based on vertex coloring, and the other one is coded multicast. 

\subsubsection{Vertex Coloring} Given the network topology $\Gc$ and the desired message pattern $\Gc_E$, we first construct the conflict graph $\Gc_c=(\Vc_c,\Ec_c)$. Every edge $(k,m)\in \Ec_E(\Gc_E)$ corresponds to a vertex $v_{km} \in \Vc_c(\Gc_c)$. That is $\Vc_c=\{v_{km}: (m,k) \in \Ec_E(\Gc_E) \}$. Two vertices $v_{km}$ and $v_{k',m'}$ are connected, i.e., $(v_{km},v_{k',m'}) \in \Ec_c(\Gc_c)$, if and only if 
\begin{itemize}
\item $k \ne k'$, indicating that two channels are not originated from the same UE, {\it and} 
\item either $(k,m') \in \Ec(\Gc)$ or $(k',m) \in \Ec(\Gc)$, indicating that (1) two channels are joint at one RRH, i.e. $m=m'$, (2) UE-$k$ interferes RRH-$m'$, or (3) UE-$k'$ interferes RRH-$m$. 
\end{itemize}
Note here that, for the conflict graph, the vertex set $\Vc_c(\Gc_c)$ is determined by the edge set $\Ec_E(\Gc_E)$, while the edge set $\Ec_c(\Gc_c)$ is determined by the edge set $\Ec(\Gc)$.

Coloring the vertices of the conflict graph ensures that the adjacent vertices (corresponding to conflicting channels) receive distinct colors (corresponding to distinct orthogonal pilots sequences). The vertices with the same color can be assigned the same pilot sequence without causing contamination in the training phase, so that the corresponding channels can be stably estimated.

\begin{example}
In Fig. \ref{fig:pilot-assign} (Right), the channels $(1,1)$, $(1,2)$ and $(1,3)$ are originated from the same UE-1, so they are not conflicting and thus can be assigned the same pilot; the channel $(1,1)$ is conflicting with all $(2,2),(2,3),(2,5),(2,7)$ and $(3,4),(4,4)$, because UE-2 interferes RRH-1 and UE-1 interferes RRH-4, respectively, if they use the same pilot. The channels $\{(1,1),(1,2),(3,3),(3,5),(4,7),(4,8)\}$ receive the same color, so that these channels can be estimated by using the same pilot sequence. The same applies to the channels $\{(2,2),(2,3),(2,5),(4,4)$, $(4,6)\}$. Thus, it can be figured out that UE-$\{1,3,4\}$ use one pilot sequence, and UE-$\{2,4\}$ use another one, and UE-4 uses the combination of those two.
\end{example}

\subsubsection{Coded Multicast} When the network topology coincides with the desired message pattern, i.e., $\Gc=\Gc_E$, meaning that all channels captured in the network topology should be estimated, we can use coded multicasting method proposed in TIM to assign pilot sequences.
Letting $T=\max_m \abs{\Tc_m}$, we can design a $(K, T)$ maximum distance separable (MDS) code with a $T \times K$ generator matrix in which any $T$ columns are linearly independent. The columns of this generator matrix can be used as pilot sequences, and each UE select one of them to use. At the RRHs, each of them obverses a combination of at most $T$ pilot signals and is able to estimate all channels. 
\begin{example}
In Fig. \ref{fig:pilot-assign} (Left), suppose all channels should be estimated. We have $\max_m \abs{\Tc_m}=3$, and hence a $(4,3)$ MDS code generator matrix can be constructed. Roughly speaking, four pilot sequences $\psiv_{1,2,3,4} \in \RR^{3 \times 1}$ are selected from the generator matrix, and any three of them are linearly independent. UE-$k$ chooses pilot sequence $\psiv_k$, and at RRH-4, the following combined pilot signal is received (with noise term omitted)
$$
\hat{\rv}_4 = g_{41} \psiv_1 + g_{43} \psiv_3 + g_{44} \psiv_4
$$
and since  $\{\psiv_t\}_{t=1,3,4}$ are linearly independent, the inverse $[\psiv_1, \psiv_3, \psiv_4]^{-1}\hat{\rv}_4$ yields the estimates of channel coefficients $\{g_{41}, g_{43}, g_{44}\}$.
\end{example}
The optimality of TIM under the groupcast setting is in general an open problem. The state-of-the-art coding schemes focuses on the information-theoretic optimality with respect to some classes of network topologies and are therefore topology-dependent. In this paper, as we are interested in the pilot assignment strategies, we aim to design achievable schemes in a systematic way although their information-theoretic optimality may be challenging to analyze.

In what follows, we first formulate a pilot assignment problem given $\Gc_E$ is known, followed by the channel pattern optimization problem with a given pilot dimension budget.

\section{Pilot Dimension Minimization}
In this section, we consider the pilot dimension minimization problem given the network topology $\Gc$ and a specified channel estimation pattern $\Gc_E$ for the uplink training.
  
Denoting by $\xv_k=[x_{k1},x_{k2},\dots,x_{kT}]^\T$, and $\Psim=[\psiv_1,\psiv_2,\dots,\psiv_T]$, we have $\sv_k=\Psim \xv_k$. 
{Each RRH-$m$ performs ``local'' interference mitigation/cancellation by combining the projections on the individual pilots $\psi_t$ and multiplying by a constant full-rank matrix $\Cm_m \in \RR^{T \times T}$.}
The resulting pilot signal $\tilde{\rv}_m=\Cm_m \hat{\rv}_{\p,m}$ can be rewritten as
\begin{align}
\tilde{\rv}_{m} &= \sqrt{\tau_\p \eta_{\p}} \Cm_m \Psim^\H \sum_{k=1}^K g_{mk}  \Psim  \xv_k +  \Cm_m \Psim^\H \wv_{\p,m} \\
&= \underbrace{\sqrt{\tau_\p \eta_{\p}}  \sum_{k: (k,m) \in \Ec_E}  \Cm_m \xv_k  g_{mk}}_{\text{desired pilot signal}} + \underbrace{\sqrt{\tau_\p \eta_{\p}}  \sum_{k: (k,m) \in \Ec \backslash \Ec_E}  \Cm_m \xv_k  g_{mk} }_{\text{significant interference}} \nonumber \\
  &\qquad+ \underbrace{\sqrt{\tau_\p \eta_{\p}}  \sum_{k: (k,m) \notin \Ec}  \Cm_m \xv_k  g_{mk}}_{\text{negligible interference}}  +  \Cm_m \Psim^\H \wv_{\p,m} \label{eq:received-pilot-signal} 
\end{align}
{where $\Cm_m$ is used to simplify problem formulation by avoiding an incomplete matrix with binary entries and will be determined later.}
It can be verified that as long as the channels are estimated from $\hat{\rv}_{\p,m}$, they can be stably estimated from $\tilde{\rv}_m$ as well with high probability.

{For a given $m$, to recover $\{g_{mk}: (k,m) \in \Ec_E, \forall k\}$ stably, 
we need to guarantee that the vectors of coefficients in $\{\Cm_m \xv_k: (k,m) \in \Ec_E, \forall k\}$ are linearly independent.}
To guarantee stable estimation, we need to let the significant interference go to zero, i.e., $\Cm_m \xv_k=0$ if $(k,m) \in \Ec \backslash \Ec_E$. The negligible interference does not contribute too much because the path loss $\beta_{mk}$ is small according to topological modeling, and therefore $\{\Cm_m \xv_k: \forall (k,m) \notin \Ec\}$ do not really matter.

In what follows, we propose a low-rank matrix completion and factorization method to calculate the minimum pilot dimension $T$ and the pilot assignment vectors $\{\xv_k\}_k$.
\subsection{Low-rank Matrix Completion and Factorization}
For the sake of problem formulation, we first construct matrix with a specific $T$ which is in fact unknown {\em a priori}, and then remove the dependence of $T$.
Collecting all vectors to form a big matrix, we have $\Cm=[\Cm_1^\T, \dots,\Cm_M^\T]^\T \in \RR^{MT \times T}$ and $\Xm=[\xv_1, \dots, \xv_K]^\T \in \{0,1\}^{K \times T}$. Let $\tilde{\Am} = \Cm \Xm^\T \in \RR^{MT \times K}$, and
$
[\tilde{\Am}]_{\tilde{\Ic}_m,k} = \Cm_{m} \xv_k \in \RR^{T \times 1}
$
where $\tilde{\Ic}_m=\{(m-1)T+1,\dots,mT\}$. Thus, the matrix form of the received pilot signal can be given by
\begin{align} \label{eq:LRMC-pilot-vector}
\tilde{\rv}_m = \sqrt{\tau_\p \eta_{\p}} \tilde{\Am}_m \gv_m + \tilde{\nv}_m
\end{align}
where $\tilde{\Am}_m = [\tilde{\Am}]_{\tilde{\Ic}_m,:}$ is the submatrix of $\tilde{\Am}$ indexed by the rows $\tilde{\Ic}_m$, and $ \tilde{\nv}_m=\Cm_m \Psim^\H \wv_{\p,m}$. Note here that, only the channels $\{g_{mk}: (k,m) \in \Ec_E\}$ are of interest to be estimated, and our goal is to figure out the matrix $\tilde{\Am}$ with rank $T$ which depends only on two patterns $\Gc$ and $\Gc_E$. 

To minimize the pilot dimension, we have
\begin{align}
T = \min \rank (\tilde{\Am})
\end{align}
where $\tilde{\Am}$ is a partially filled matrix and is supposed to possess the following property:
\begin{align}
[\tilde{\Am}]_{\tilde{\Ic}_m,k} = \left \{ \Pmatrix{\tilde{\cv}_{mk}, & \text{if } (k,m) \in \Ec_E \\ \mathbf{0}, & \text{if } (k,m) \in \Ec \backslash \Ec_E \\  *, & \text{otherwise}} \right.
\end{align}
where $\tilde{\cv}_{mk}$ is any nonzero vector, and $*$ is any indefinite $T \times 1$ vector. To ensure that the channels of interest $\{g_{mk},  (k,m) \in \Ec_E\}$ can be stably estimated over $T$ pilot dimensions, the following should be satisfied:
\begin{align} \label{eq:rank-constraint}
\rank ([\tilde{\Am}]_{\tilde{\Ic}_m, \Tc_{E,m}}) = \abs{\Tc_{E,m}}.
\end{align}
For simplicity, $[\tilde{\Am}]_{\tilde{\Ic}_m,\Tc_{E,m}}$ can be chosen from the columns of the identity matrix $\Id_{T}$.

Observing that each RRH is not connected to all UEs, we note that some rows in $\tilde{\Am}$ may only have zero or indefinite elements. The rank minimization is prone to turning these rows to be all zero, i.e., by setting indefinite elements to be 0. As such, we can safely remove these rows from $\tilde{\Am}$ without reducing the rank. Because RRH-$m$ has $\abs{\Tc_m}$ connected UEs, so there are $\abs{\Tc_m}$ nonzero vectors with a single nonzero element in $\{[\tilde{\Am}]_{m,1},\dots,[\tilde{\Am}]_{m,K}\}$ and the rest is indefinite. By this, we only need to keep the  $\abs{\Tc_m}$ rows with nonzero elements in $[[\tilde{\Am}]_{m,1},\dots,[\tilde{\Am}]_{m,K}]$. In doing so, a modified matrix $\Am$ has in total $\sum_{m=1}^M\abs{\Tc_m}$ rows and possesses the following property:
\begin{align} \label{eq:matrix-structure}
[{\Am}]_{\Ic_m,k} = \left \{ \Pmatrix{{\cv}_{mk}, & \text{if } (k,m) \in \Ec_E \\ \mathbf{0}, & \text{if } (k,m) \in \Ec \backslash \Ec_E \\  *, & \text{otherwise}} \right.
\end{align}
where $\cv_{mk}$ can be any $\abs{\Tc_m} \times 1$ vector, ${\Ic}_m=\{\sum_{m'=1}^{m-1}\abs{\Tc_{m'}}+1,\dots,\sum_{m'=1}^m\abs{\Tc_{m'}}\}$, and the full column rank property of $[\Am]_{\Ic_m, \Tc_{E,m}}$ should be maintained.
Thus, we have the low-rank  matrix completion problem formulation
\begin{subequations} \label{eq:low-rank-MC}
\begin{align}
T = \min_{\Am}  & \quad \rank (\Am)\\
\st  & \quad \rank([\Am]_{\Ic_m, \Tc_{E,m}}) = \abs{\Tc_{E,m}}, \quad \forall m.
\end{align}
\end{subequations}
where $\Am$ follows the structure in \eqref{eq:matrix-structure}.
This matrix completion problem is known to be difficult to solve. 
Instead of pursuing the unique completion as in the literature, we are only interested in finding one feasible solution with any properly filled indefinite entries.
Thus, for a given rank $r$, we reformulate this problem as a feasibility problem as follows 
\begin{align} \label{eq:low-rank-MC-feasibility}
\text{find} \quad \Am, \qquad
\st \quad \rank (\Am) \le r, \;
 \rank([\Am]_{\Ic_m, \Tc_{E,m}}) = \abs{\Tc_{E,m}}, \; \forall m.
\end{align}

Thus, denoting by $\bar{M}=\sum_{m=1}^M \abs{\Tc_m}$, $r_m=\abs{\Tc_{E,m}}$, and $\Ic'_m=\{\sum_{m'=1}^m \abs{\Tc_{m'}}+1:\sum_{m'=1}^m \abs{\Tc_{m'}}+r_m\}$, we define three constraint sets:
\begin{subequations} \label{eq:constraint-sets}
\begin{align}
\Sc_{\Omega} &= \{ \Am \in \RR^{\bar{M} \times K}: [{\Am}]_{\Omega} = \mathbf{0} \}\\
\Sc_r &= \{ \Am \in \RR^{\bar{M} \times K}: \rank (\Am) \le r\} \\
\Sc_{\Omega_E} &= \{ \Am \in \RR^{\bar{M} \times K}: [\Am]_{\Ic'_m, \Tc_{E,m}} = \Id_{r_m}, \forall m \}
\end{align}
\end{subequations}
where $\Omega = \{(\Ic_m, k): (k,m) \in \Ec \backslash \Ec_E\}$, $\Omega_E = \{(\Ic_m, k): (k,m) \in \Ec_E\}$, and ${\ev}_{mk}$ is $k$-th column of the identity matrix $\Id_{\abs{\Tc_m}}$.

Such a low-rank matrix completion formulation is a generalized version of that for the multiple-unicast TIM problem \cite{Hassibi,TIM-LRMC}. In a similar way, we can adopt a low-complexity alternating projection method \cite{Hassibi} to obtain a feasible solution (see Alg. \ref{alg:low-rank}) by projecting iteratively on the above constraint sets, e.g., $\Pc_{\Sc}(\Am)$ is to project $\Am$ onto the set $\Sc$.
\begin{algorithm}
\caption{Matrix Completion via Alternating Projection}
\label{alg:low-rank}
{\bf Input:}  $\Gc$, $\Gc_E$.
\begin{algorithmic}[1]
\For{$r=K, K-1, \dots, 1$}
\State Set $k=0$, and randomly generate $\Am_r^0, \Bm_r^0 \in \RR^{\bar{M} \times K}$
\While{$\Norm{\Am_r^k-\Bm_r^k} > \epsilon$ \& $k \le \text{It}_{\max}$ }
\State $\Bm_r^k \gets \Pc_{\Sc_r}(\Am_r^k)$
\State $\Am_r^{k} \gets \Pc_{\Sc_\Omega}(\Bm_r^k) + \Pc_{\Sc_{\Omega_E}}(\Bm_r^k)$
\State $k \gets k+1$
\EndWhile
\State \textbf{If} {$k < \text{It}_{\max}$} \textbf{then}
 Update $\Am \gets \Am_{r}^{k}$ and break 
\textbf{end if}
\EndFor
\end{algorithmic}
{\bf Output:} $T=r$, $\Am$.
\end{algorithm}
\vspace{-5pt}

Once $\Am$ is completed, inserting zero rows gives us the original matrix $\tilde{\Am}$. Then the matrix  $\tilde{\Am}$ will be factorized into a real matrix $\Cm$ and a binary matrix $\Xm$, i.e., $\tilde{\Am} = \Cm \Xm^\T$ where $\Cm \in \mathbb{R}^{MT \times T}$ and $\Xm \in \{0,1\}^{K \times T}$.
This is a matrix factorization problem with binary component that arises in various problems, such as blind binary source signal separation and network component analysis.
Although no existing algorithms guarantee the exact unique factorization due to the non-convexity, some efficient algorithms were proposed to yield a feasible solution. The problem can be efficiently done by adopting the low-complexity algorithm in \cite{BMF}, by which we obtain a feasible pilot assignment $\xv_k=\Xm_k$ for all $k$.  
Once the pilot assignment $\{\xv_k\}_k$ is determined, the MMSE channel estimator as in \eqref{eq:channel-estimate} can be applied to produce channel estimates $\{\hat{g}_{mk}\}_{m,k}$. 

\section{Channel Pattern Optimization}
When the channel estimation pattern is unknown {\em a priori}, the pilot assignment needs to be done together with the optimization of such a pattern. In what follows, we consider the pilot assignment problem given a budget of pilot dimension $T$ when $\Gc_E$ is unknown.

We take a closer look at each pilot assignment indicators $\{x_{kt}\}$, enforcing that each pilot should be used to estimate at most one channel at each RRH.
To this end, we introduce another set of binary variables $\{y_{mt}\}$ such that
\begin{align}
y_{mt} = \left\{\Pmatrix{1, & \text{if RRH-$m$ estimates using $\psiv_t$ with success,}\\0, & \text{otherwise,}}\right.
\end{align}
where $y_{mt}$ indicates whether or not the pilot $\psiv_t$ is useful for channel estimation. In terms of success, we meant the channel between RRH-$m$ and UE-$k$ can be stably estimated when UE-$k$ is assigned with the pilot $\psiv_t$ and RRH-$m$ is using the same pilot $\psiv_t$.

{We further assume that each pilot $\psiv_t$ at RRH-$m$ can at most estimate channels from $\kappa$ UEs connected to RRH-$m$ by e.g., zero-forcing. Thus, we have the following constraint
\begin{align} \label{one-estimated}
\sum_{k \in \Tc_m} x_{kt} y_{mt} \le \kappa, \quad \forall m,t
\end{align}
where $\kappa=1$ means RRH-$m$ is dedicated to one single UE for pilot $\psiv_t$.
}

For ease of presentation, we define a topology matrix $\Tm \in \{0,1\}^{K \times M}$ as follows:
\begin{align}
[\Tm]_{km} = \left \{\Pmatrix{1, &\text{if $(k,m) \in \Ec(\Gc)$}\\ 0, &\text{otherwise.}} \right.
\end{align}
Given the budget of pilot dimension $T$, the objective of pilot assignment is to make sure that as many strong channels as possible can be stably estimated by pilot $\{\psiv_t\}_{t=1}^T$. That is, 
\begin{subequations}
\begin{align}
\max_{\{x_{kt},y_{mt}\}}  \quad &  \sum_{t=1}^T \sum_{m=1}^M \sum_{k=1}^K   [\Bm_T]_{km} x_{kt} y_{mt}\\
\st \quad 
& \sum_{k=1}^K  [\Tm]_{km} x_{kt} y_{mt} \le \kappa, \quad \forall m,t \\
& x_{kt}, y_{mt} \in \{0,1\}, \quad \forall k, m, t
\end{align}
\end{subequations}
where $[\Bm_T]_{km}=\beta_{km}$, and the objective is to find a set of triples $(m,k,t)$ with maximum sum weights $\{\beta_{mk}\}$. For any given $(m, t)$, the selected triples are subject to the constraint \eqref{one-estimated}.

\subsection{Binary Quadratically Constrained Quadratic Programming}

The above optimization problem can be rewritten in a matrix form as
\begin{subequations}
\begin{align}
\max_{\Xm, \Ym}  \quad &  \vv_X^\T \Qm_0 \vv_Y\\
\st \quad 
& \vv_X^\T \Qm_{m,t} \vv_Y \le \kappa, \quad \forall m,t \\
& \vv_X \in \{0,1\}^{KT}, \vv_Y \in \{0,1\}^{M  T}
\end{align}
\end{subequations}
where $\vv_X=\Vec(\Xm)$ and $\vv_Y=\Vec(\Ym)$ are vectorization of the corresponding matrices, and
\begin{align}
\Qm_0 &=(\Bm_T \odot \Tm) \otimes \Id_T\\
\Qm_{m,t} &= (\Tm \odot (\one_K \otimes \ev_m^\T)) \otimes \diag (\ev_t)
\end{align}
in which $\one_K$ is the $K \times 1$ all-one vector, $\ev_m$ is the $m$-th column of $\Id_M$, and $\ev_t$ is the $t$-th column of $\Id_T$.
This is a binary quadratically constrained quadratic program (BQCQP), in which two set of binary parameters $\{x_{kt}\}$ and $\{y_{mt}\}$ are interacting each other. This type of problems is known to be difficult to solve. A possible approach is to relax the BQCQP problem by SDP relaxation as in \cite{BQCQP}.

\subsection{Sequential Maximum Weight Induced Matching (sMWIM)}
A more tractable solution is to consider each pilot sequentially, so that for each pilot, we assign it to as many UE-RRH links as possible, and after $T$ sequential assignment, the resulting assignments are expected to achieve a good approximation of the original problem. 

First, let us focus on the pilot assignment for a given pilot sequence $\psiv_t$ and a given network topology $\Gc$. The goal is to assign the same pilot to as many UE-RRH links as possible. The optimization subproblem can be formulated as follows:
\begin{subequations} \label{MWIM}
\begin{align} 
\max  \quad & \sum_{m=1}^M \sum_{k=1}^K  [\Bm_T]_{km} x_{kt} y_{mt}\\
\st \quad & x_{kt} \le \sum_{m=1}^M [\Tm]_{km} y_{mt}, \quad \forall k \label{RRH-constraint}\\
&  y_{mt} \le  \sum_{k=1}^K [\Tm]_{km} x_{kt}, \quad \forall m \label{UE-constraint} \\
& \sum_{k=1}^K [\Tm]_{km} x_{kt} \le \kappa y_{mt} + K(1-y_{mt}), \quad  \forall m\label{induced-matching-constraint}\\
& x_{kt}, y_{mt} \in \{0,1\}, \quad \forall k, m \label{binary-variable}
\end{align}
\end{subequations}
where \eqref{RRH-constraint} indicates that if UE-$k$ is assigned the pilot $\psiv_t$, then there is at least one RRH with strong connections to UE-$k$ is able to estimate the channel coefficient by using the pilot $\psiv_t$; \eqref{UE-constraint} indicates that if an RRH can estimate the channel coefficient using pilot $\psiv_t$, then there is at least one UE sending such a pilot; and \eqref{induced-matching-constraint} guarantees that if the RRH-$m$ can estimate the channel coefficient using the pilot $\psiv_t$, {there exist at most $\kappa$ UEs with strong connectivity to this RRH that can be assigned with this pilot.} These constraints are to ensure that of \eqref{one-estimated}. Note that there is not a similar constraint of  \eqref{induced-matching-constraint} for UEs, meaning that one UE can use the same pilot to train multiple channels as long as the RRHs are capable to do so.

This can be recognized as a modified version of the classic maximum weight induced matching problem in a quadratic programming form. Here the difference from the conventional induced matching is that, (1) there may exist multiple edges originated from the same $k$, corresponding to the scenario that the channel coefficients from a UE to multiple RRHs can be estimated at these RRHs using the same pilot; (2) there exist multiple edges from the same $m$, meaning that channels from multiple UEs can be estimated at the same RRH.

Let us linearize it into the following form by introducing an auxiliary variable $z_{mkt}=x_{kt}y_{mt}$:
\begin{subequations} \label{linearized}
\begin{align}
\max  \quad & \sum_{m=1}^M \sum_{k=1}^K \beta_{mk}  z_{mkt} \label{eq:obj-func}\\
\st \quad & \eqref{RRH-constraint} - \eqref{binary-variable}\\
& z_{mkt} \le x_{kt}, \quad \forall (k,m) \in \Ec \label{eq:cons-z1}\\
& z_{mkt} \le y_{mt}, \quad \forall (k,m) \in \Ec\\
&z_{mkt} \ge x_{kt} + y_{mt} - 1, \quad \forall (k,m) \in \Ec\\
& z_{mkt} \in \{0,1\}, \quad \forall m,k \label{eq:cons-z4}
\end{align}
\end{subequations}
where these additional constraint is to ensure that $z_{mkt}=1$ if and only if $x_{kt}=y_{mt}=1$.
In general, this optimization is a linear integer program, and can be solved by applying off-the-shelf solvers. Taking a closer look at the additional constraints for $\{z_{mkt}\}$, we observe that \eqref{eq:cons-z4} can be relaxed without loss of optimality, that is, $z_{mkt} \in \{0,1\}$ can be relaxed to $z_{mkt} \in [0,1]$, owing to the integer-valued $\{x_{kt}\}_k$ and $\{y_{mt}\}_m$. For a large-scale network with large $M$ and $K$, as the computational complexity of \eqref{linearized} is still prohibitively high, we can follow Benders' decomposition in \cite{ahat2017integer} to separate the variables $\{x_{kt},y_{mt}\}$ from $\{z_{mkt}\}$ to reduce complexity.

Then, we can sequentially solve \eqref{linearized} with reweighed $\beta_{km}$, so that the assigned UE-RRH link $(k,m)$ will not be reconsidered later. 
Benders' Decomposition is to first search for a feasible induced matching by optimizing a master problem with variables $\{x_{kt},y_{mt}\}_{k,m}$ and the constraints \eqref{RRH-constraint} - \eqref{binary-variable}, followed by a slave subproblem to maximize the objective function \eqref{eq:obj-func} with variables $\{z_{mkt}\}_{k,m}$ and the constraints \eqref{eq:cons-z1}-\eqref{eq:cons-z4}. The master and slaver problems will be connected with a refined cut as defined below.
Specifically, in order not to select the same set of edges as induced matching for different pilot dimension, we introduce $\Tm^{(t)}$ to denote the remaining network topology with the selected edges in the previous pilot dimensions removed, where $\Tm^{(0)}$ represent the initial network topology $\Gc$. Thus, the master problem turns out to be
\begin{subequations} \label{master}
\begin{align} 
\max  \quad & \sum_{m=1}^M  y_{mt} + L\\
\st \quad & x_{kt} \le \sum_{m=1}^M [\Tm^{(t)}]_{km} y_{mt}, \quad \forall k \label{master:RRH-constraint}\\
&  y_{mt} \le  \sum_{k=1}^K [\Tm^{(t)}]_{km} x_{kt}, \quad \forall m \label{master:UE-constraint} \\
& \sum_{k=1}^K [\Tm^{(0)}]_{km} x_{kt} \le \kappa y_{mt} + K(1-y_{mt}), \quad  \forall m\label{master:induced-matching-constraint}\\
& L \le \sum_{m=1}^M \sum_{k=1}^K \hat{L}^*(x_{kt},y_{mt}) \label{eq:benders-cut} \\
& x_{kt}, y_{mt} \in \{0,1\}, \quad \forall k, m, 
\end{align}
\end{subequations}
where \eqref{eq:benders-cut} is the Benders' cut that will be determined later.
Denote by $(\{\hat{x}_{kt}\}_k, \{\hat{y}_{mt}\}_m,\hat{L})$ the optimal solution to the master problem. The slave problem can be given by
\begin{subequations} \label{slave}
\begin{align}
\max  \quad & \sum_{m=1}^M \sum_{k=1}^K [\Bm^{(t)}_T]_{km}  z_{mkt} \label{slave:obj-func}\\
\st \quad & z_{mkt} \le \hat{x}_{kt}, \quad \forall (k,m) \in \Ec\\
& z_{mkt} \le \hat{y}_{mt}, \quad \forall (k,m) \in \Ec\\
&z_{mkt} \ge \hat{x}_{kt} + \hat{y}_{mt} - 1, \quad \forall (k,m) \in \Ec\\
& z_{mkt}\ge 0, \quad \forall m,k
\end{align}
\end{subequations}
whose dual problem can be given by
\begin{subequations} \label{slave-dual}
\begin{align}
\min_{\{a_{km}, b_{km}, c_{km}\}} & \sum_{m=1}^M \sum_{k=1}^K \big(a_{km} \hat{x}_{kt} + b_{km} \hat{y}_{mt} + c_{km} (\hat{x}_{kt}+\hat{y}_{mt}-1) \big) \\
\st \quad & a_{km} + b_{km} + c_{km} \ge [\Bm^{(t)}_T]_{km}, \quad \forall k, m\\
& a_{km} \ge 0, b_{km} \ge 0, c_{km} \le 0, \quad \forall (k,m) \in \Ec.
\end{align}
\end{subequations}
Let the optimal solution to \eqref{slave-dual} be $\{\hat{a}_{km}, \hat{b}_{km}, \hat{c}_{km}\}$. The updated Benders' cut can be refined by
\begin{align} \label{eq:upper-bound}
\hat{L}^*(x_{kt},y_{mt}) = \hat{a}_{km} x_{kt} + \hat{b}_{km} y_{mt} + \hat{c}_{km} (x_{kt}+y_{mt}-1).
\end{align}

The sMWIM algorithm is summarized in Alg.~\ref{alg:smwim}. It has a multi-round procedure. In each round $t$, we find the maximum weight induced matching over the remaining network topology $\Tm^{(t)}$, by solving both the master and slave problems \eqref{master}-\eqref{slave} iteratively, until the update of Benders' cut stabilizes. The algorithm continues until $t$ exceeds the maximum pilot dimension $T_{\max}$ or all edges in $\Gc$ are assigned with a pilot. It is worth noting that the approach assigns orthogonal pilots to each UE-RRH link individually, such that one UE may be assigned with the combination of multiple pilots finally, each of which is dedicated to some UE-RRH links.

\begin{algorithm}
\caption{Sequential Maximum Weight Induced Matching (sMWIM)}
\label{alg:smwim}
{\bf Input:} $\Tm$, $\Bm_T$, $T_{\max}$, $\kappa$.
\begin{algorithmic}[1]
\State {\bf Initialization:} $\Tm^{(1)}=\Tm$, $\Bm_T^{(1)}=\Bm_T$, $t=1$
\While {$t \le T_{\max} \; \& \; \Tm^{(t)} > 0$ } 
\State Set $j=1$, $L_1^*(t)=\norm{\Bm_T^{(t)}}_1$, $L_0^*(t)=0$
\While {$\abs{L_j^*(t)-L_{j-1}^*(t)} > \epsilon$}
\State Solve \eqref{master} and obtain $\{{x}_{kt}\}_k$ and $\{{y}_{mt}\}_m$
\State Solve \eqref{slave} and obtain $\{{z}_{mkt}\}_{k,m}$
\State Update $L_{j+1}^*(t) \gets \hat{L}^*({x}_{kt}, {y}_{mt})$ according to \eqref{eq:upper-bound}
\State Update $j \gets j+1$
\EndWhile
\State Update $[\Tm^{(t+1)}]_{km} \gets [\Tm^{(t)}]_{km} - {z}_{mkt}$, for all $k,m$
\State Update $\Bm_T^{(t+1)} \gets \Bm_T^{(t)} \odot \Tm^{(t+1)}$
\State Update $t \gets t+1$;
\EndWhile
\end{algorithmic}
{\bf Output:} $\{x_{kt}\}_{k,t}$, $\{y_{mt}\}_{m,t}$, $T= t-1$.
\end{algorithm}

\subsection{Greedy Algorithm}
While the sMWIM algorithm gives us a tractable solution, the computational complexity of the mixed integer program formulation usually scales with the number of parameters, even if Benders' decomposition is applied. By revisiting the formulation in \eqref{MWIM}, we take a step back to formulate the TPA problem as a many-to-many matching problem instead of the induced matching, for which we develop a greedy algorithm to find a feasible solution.

By letting $z_{mkt}=x_{kt}y_{mt}$, for the $t$-th round, the optimization \eqref{MWIM} can be replaced by a many-to-may matching problem with the following linear integer program formulation
\begin{subequations}\label{eq:GMAP}
\begin{align}
\max \quad &  \sum_{m=1}^{M}\sum_{k=1}^K\left[\tilde{\Bm}_T^{(t)}\right]_{km}z_{mkt}\label{eq:GMAP-1}\\
\mathrm{s.t.} \quad & \sum_{k=1}^K z_{mkt}\leq \kappa, \quad \forall m \label{eq:GMAP-2}\\
&\sum_{m=1}^M z_{mkt}\leq \kappa_u, \quad \forall k \label{eq:GMAP-3}\\
&z_{mkt}\in\{0,1\}, \quad \forall m, k\label{eq:GMAP-4}
\end{align}
\end{subequations}
where $\kappa$ and $\kappa_u$ denote the maximum number of UEs that each RRH could serve and the number of connected RRHs per user, respectively. For simplicity, we set $\kappa$ and $\kappa_u$ as constant integers throughout the iteration. 
The above many-to-many matching problem is also known as the generalized multi-assignment problem (GMAP) \cite{park1998lagrangian}.

\begin{algorithm}
\caption{TPA via Greedy Algorithm}
{\bf Input}: $\Tm$, $\Bm_T$, $T_{\max}$, $\kappa$, $\kappa_u$
\begin{algorithmic}[1]
\State {\bf Initialization}: ${\Tm}^{(1)}= \Tm$, ${\Bm}_T^{(1)}= \Bm_T$, $t=1$
\While{$t \le T_{\max} \; \& \; {\Tm}^{(t)} > 0$}
\State Set $\text{FLAG}=1$, $\tilde{\Tm}^{(t)}={\Tm}^{(t)}$, $\tilde{\Bm}_T^{(t)}={\Bm}_T^{(t)}$, $x_{kt}=y_{mt}=1$ for all $k \in [K], m \in [M]$
\State Update $\tilde{\Tm}_{\max}^{(t)}$ and $\tilde{\Bm}_{T,\max}^{(t)}$ according to \eqref{T_max}
\For{$m \in \{m': \sum_{k=1}^{K} [\tilde{\Tm}_{\max}^{(t)}]_{km'}> 1, \forall m' \in [M]\}$}
\State Update $\tilde{\Tm}^{(t)}$ such that $[\tilde{\Tm}^{(t)}]_{k,:}=\zerov, \; \forall k\notin \arg\max_{i}\{[\tilde{\Bm}_T^{(t)}]_{im}\}$
\State Update $\tilde{\Bm}_T^{(t)}\gets \tilde{\Bm}_T^{(t)}\odot \tilde{\Tm}^{(t)}$
\EndFor
\For{$k \in \{k': \sum_{m=1}^{M} y_{mt}[\tilde{\Tm}^{(t)}]_{k'm}> \kappa_u, \forall k' \in [K]\}$}
\State Update $\tilde{\Tm}^{(t)}$ such that $[\tilde{\Tm}^{(t)}]_{km}=0, \; \forall m\notin \arg\max^{\kappa_u}\{[\tilde{\Bm}_T^{(t)}]_{km'}\}$
\EndFor
\State Define profit and cost matrices $\Pm^{(t)}$ and $\Cm^{(t)}$ as \eqref{profit} and \eqref{cost}
\While{\text{FLAG}}  
\State Select the RRH $m$ such that $\sum_{k=1}^{K} x_{kt} [\tilde{\Tm}^{(t)}]_{mk} > \kappa$ 
\State Compute \eqref{final-profit} as $\Phi^b$ if the RRH-$m$ is not selected, i.e., $x_{kt}=0$
\State Compute \eqref{final-profit} as $\Phi^u$ if only $\kappa$ UEs with largest elements in $\tilde{\Bm}_T^{(t)}$ are selected 
\If{$\Phi^b > \Phi^u$}
\State $y_{mt} = 0$, and $[\tilde{\Tm}^{(t)}]_{km}=0$, $\forall k\in[K]$
\Else
\State $x_{kt}=0$, and $[\tilde{\Tm}^{(t)}]_{km}=0, \forall m\in[M]$, $k\notin \max^{\kappa}\{i:[\tilde{\Bm}_T^{(t)}]_{im},i\in [K]\}$
\EndIf
\State Update $\tilde{\Bm}_T^{(t)}\gets \tilde{\Bm}_T^{(t)}\odot \tilde{\Tm}^{(t)}$
\If {$\sum_{k=1}^{K}{x_{kt}[\tilde{\Tm}^{(t)}]_{km}}\leq \kappa, \forall m \in [M]$}
\State $\text{FLAG} = 0$
\EndIf
\EndWhile
\State Update $[{\Tm}^{(t+1)}]_{km} \gets [{\Tm}^{(t)}]_{km} - x_{kt}$, $\forall k,m$
\State Update ${\Bm}_T^{(t+1)}\gets {\Bm}_T^{(t)}\odot {\Tm}^{(t+1)}$
\EndWhile
\end{algorithmic}  
{\bf Output}: $\{x_{kt}\}_{k,t}$, $\{y_{mt}\}_{m,t}$ 
\label{greedy}
\end{algorithm}

To solve the GMAP in an efficient way, we develop a greedy algorithm as shown in Alg. \ref{greedy}. 
Given the initial network topology $\Gc$, which can be constructed with or without RRH selection, we take at most $T_{\max}$ rounds to assign pilot sequences to different UEs. At the $t$-th round, we introduce an auxiliary adjacency matrices $\tilde{\Tm}^{(t)}$ (and the corresponding path loss matrices $\tilde{\Bm}_T^{(t)}$) to indicate the remaining network topology to be considered for pilot assignment. Once the UEs are assigned with pilots, they will be removed from consideration, which yields an updated $\tilde{\Tm}^{(t+1)}$ (see Line 27 in Alg. \ref{greedy}). It is worth pointing out that $\tilde{\Tm}^{(t)}$ is usually not equal to ${\Tm}^{(t)}$ in the previous section, because of the use of different matching algorithms.

At the $t$-th round, we have a pre-selection procedure to identify the network topology $\tilde{\Tm}^{(t)}$ for the many-to-many matching. First, we introduce a binary matrix $\tilde{\Tm}_{\max}^{(t)}$ to indicate the position of the maximum coefficients, defined as 
\begin{align}
\label{T_max}
[\tilde{\Tm}_{\max}^{(t)}]_{km} = \left\{ \Pmatrix{1, \quad \text{if $[\tilde{\Bm}_{T}^{(t)}]_{km}= \max_{m} \{[\tilde{\Bm}_{T}^{(t)}]_{km}\}$,}\\ 0, \quad \text{otherwise,}} \right.
\end{align}
and the corresponding path loss matrix $\tilde{\Bm}_{T,\max}^{(t)}=\tilde{\Bm}_{T}^{(t)} \odot \tilde{\Tm}_{\max}^{(t)}$ to pre-select the maximum coefficients. In each round, if there are multiple UEs that compete for the same RRH, then only the one with the largest path loss coefficient will be considered in this round, and the rows corresponding to other competing UEs in $\tilde{\Bm}_T^{(t)}$ will be set to zero (see Lines 5-8 in Alg. \ref{greedy}). In doing so, we try to ensure each UE can be served by the dominant RRH with the largest path loss coefficient and avoid the competition for the dominant RRH between UEs in the same round. Second, for the selected UEs, if the number of connected active RRHs is larger than $\kappa_u$, then only the RRHs with the largest $\kappa_u$ path loss coefficients will be considered, and others will be removed from the topology (see Lines 9-11 in Alg. \ref{greedy}, where $\max^{p} \Ac$ is to choose the largest $p$ elements from $\Ac$). By doing so, the constraint \eqref{eq:GMAP-3} is automatically satisfied. 
Third, we select RRHs that do not satisfy the constraint \eqref{eq:GMAP-2} and make the decision to either switch off these RRHs (i.e., $y_{mt}=0$) or some UEs (i.e., $x_{kt}=0$) to make \eqref{eq:GMAP-2} satisfied (see Lines 14-21 in Alg. \ref{greedy}). To make the decision, we introduce the following evaluation function for the $t$-th round 
\begin{align}
\Phi^{(t)} = \sum_{m=1}^M\sum_{k=1}^K x_{kt} y_{mt}\left([\Pm^{(t)}]_{km}-\delta[\Cm^{(t)}]_{km}\right)\label{final-profit}
\end{align}
where $\delta$ is a predefined parameter to compromise between profit and cost, defined as
\begin{align}
&[\Pm^{(t)}]_{km} = \sum_{j=1}^K [\tilde{\Bm}_T^{(t)}]_{km}[\tilde{\Bm}_T^{(t)}]_{jm},\label{profit}\\
&[\Cm^{(t)}]_{km} = \sum_{j=1,j\neq k}^K [\tilde{\Bm}_T^{(t)}]_{km}[\tilde{\Bm}_T^{(t)}]_{jm},\label{cost}
\end{align}
for all $k,m$.
It is worth noting that both profit and cost matrices rely only on the path loss information $\{\beta_{mk}\}_{m,k}$ for assignment, which is different from the existing approaches in the literature.
A similar approach was also demonstrated to be effective and efficient in active channel sparsification in FDD massive MIMO systems \cite{yu2021downlink}.

\section{Numerical Results}
\label{sec:sim}
In this section, we evaluate our proposed TPA algorithms via simulations under the cell-free massive MIMO settings \cite{Cell-Free}.
We consider a square area of 1 km $\times$ 1 km in the dense urban scenario where $M$ RRHs and $K$ UEs with single antenna are uniformly located at random. To avoid the boundary effects, we also let the area be wrapped around for the random placement of RRHs. The large-scale fading coefficient $\beta_{mk}$ is modeled as follows:
\begin{align}
10 \log_{10} (\beta_{mk}) = {\rm PL}_{mk} + \sigma_{\rm sh} n_{mk}
\end{align}
where $ {\rm PL}_{mk}$ represents the path loss (in dB) between RRH-$m$ and UE-$k$, and $\sigma_{\rm sh}$ denotes the standard deviation (in dB) of shadow fading with $n_{mk} \sim \CN(0,1)$. We mainly focus on the uncorrelated shadowing model for simplicity.
In our simulation, a three-slope path loss model \cite{Cell-Free} is considered. Specifically,
\begin{align}
{\rm PL}_{mk} = \left\{ \Pmatrix{ -L-15\log_{10}(d_1)-20\log_{10}(d_0),  \hspace{1cm} \text{if $d_{mk} \le d_0$}\\ -L-15 \log_{10}(d_1)-20\log_{10}(d_{mk}), \qquad \text{if $d_0<d_{mk}\le d_1$} \\ -L-35 \log_{10}(d_{mk}), \hspace{3.5cm} \text{if $d_{mk}>d_1$}} \right.
\end{align}
where $d_{mk}$ is the distance (m) between RRH-$m$ and UE-$k$, and we use Hata-COST231 propagation model when $d_{mk}>d_1$ with $d_0=10$ m and $d_1=50$ m. Here, we have
\begin{align*}
\MoveEqLeft L \defeq 46.3+333.9 \log_{10}(f)-13.82\log_{10}(h_{a}) - (1.1\log_{10}(f)-0.7)h_u + (1.56\log_{10}(f)-0.8)
\end{align*}
where $f$ is the carrier frequency (MHz), and $h_a$ and $h_u$ are the heights (m) of RRHs and UEs, respectively. The values of these system parameters are summarized in Table \ref{tab:param}.
The following baseline pilot assignment algorithms are chosen for comparison.
\begin{itemize}
\item \textbf{Semi-random \cite{Cell-Free}}: Each user randomly chooses one orthongonal pilot, so that for each pilot dimension, $\lceil \frac{K}{T} \rceil$ users are randomly selected. 
\item \textbf{Cell-free greedy \cite{Cell-Free}}: $K$ users are assigned with $K$ pilots randomly, and the users with low downlink rate will be iteratively reassigned with new pilots to minimize pilot contamination.
\item \textbf{Structured policies \cite{chen2020structured}}: The user group scheme with RRH selection is adopted. This is a state-of-the-art pilot assignment method for cell-free massive MIMO.
\item \textbf{TPA LRMC+Semi-random}: Alg. \ref{alg:low-rank} is applied to obtain the minimal dimension of the required pilots, and the semi-random method is adopted for pilot assignment. 
\item \textbf{TPA sMWIM}: Alg. \ref{alg:smwim} is applied to find the set of binary values $\{x_{kt}\}_{k,t}$ such that the pilot $\psiv_t$ will be assigned to the user $k$ when $x_{kt}=1$. 
\item \textbf{TPA greedy}: Alg. \ref{greedy} is applied to find $\{x_{kt}\}_{k,t}$ such that the pilot $\psiv_t$ will be assigned to the user $k$ when $x_{kt}=1$.
\end{itemize}

Unless otherwise explicitly specified, we consider $M=100$, $K=40$, and $\kappa_u=20$ in our simulations. Table \ref{tab:param} lists all system parameters. We use $\Gc=30\%$ to indicate that $30\%$ of UE-RRH links with largest $\{\beta_{mk}\}$ will be considered for channel estimation.

\begin{table}
\center
\caption{System Parameters}
  \begin{tabular}{ c | c }
    \hline \hline
    Parameters & Values\\ \hline
    Cell range & 1km $\times$ 1km \\ \hline
    Carrier Frequency & 1900 MHz \\ \hline
    Bandwidth & 20 MHz \\ \hline
    Power $\rho_{\p}$~/~$\rho_{\dd}$ & 100mW/200mW \\ \hline
    Noise power spectral density & -174 dBm/Hz \\ \hline
    Antenna Height RRH/UE & 15m/1.65m \\ \hline
    Shadow Fading $\sigma_{\rm sh}$ & 8 dB \\ \hline
    Noise Figure & 9 dB \\ \hline \hline
  \end{tabular}
  \label{tab:param}
  \end{table}
  
 \begin{figure}
 \begin{minipage}{0.48\linewidth}
 \centering
\includegraphics[width=1\columnwidth]{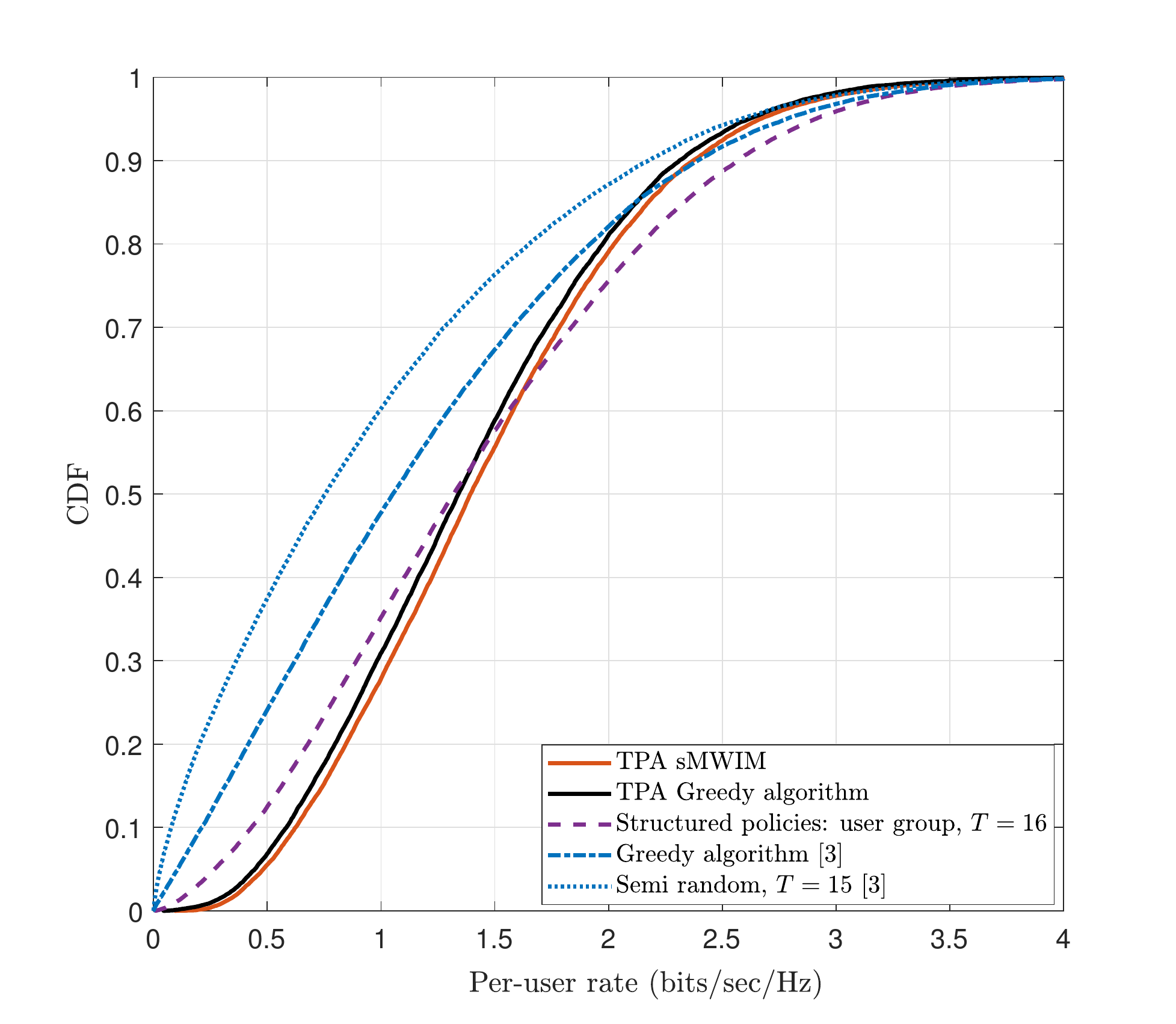}
\caption{The CDF of the downlink achievable rate per user with $\Gc=75\%$ and $\kappa=2$.}
\label{fig:70p_30p_CDF}
\end{minipage}
\quad
\begin{minipage}{0.48\linewidth}
\includegraphics[width=1\columnwidth]{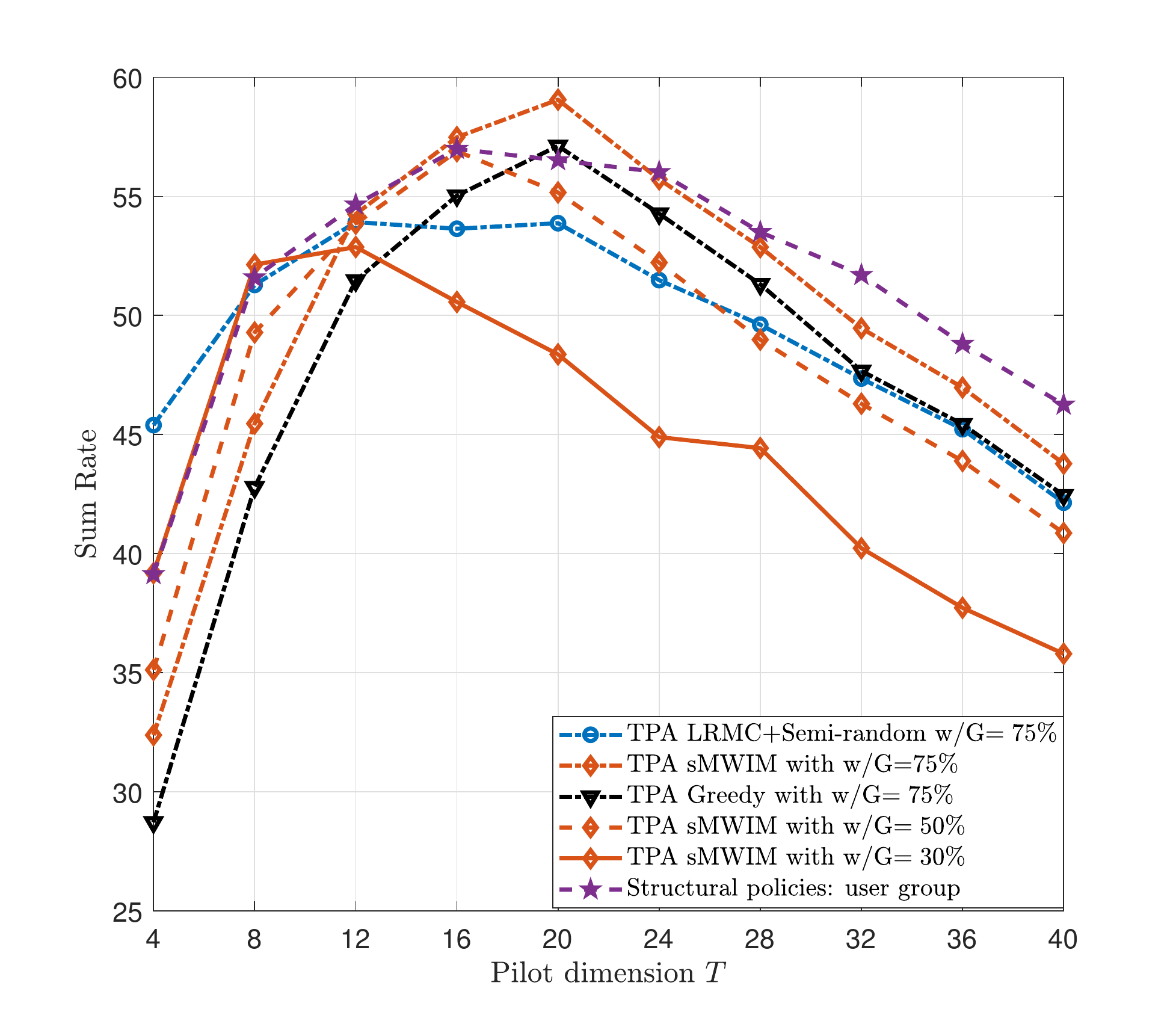}
\caption{The downlink achievable sum rate versus pilot dimension $T$.}
\label{fig:50p_30p_CDF}
\end{minipage}
\vspace{-20pt}
\end{figure}

In Figure \ref{fig:70p_30p_CDF}, we compare the cumulative distribution function (CDF) of the downlink achievable rate per user of our proposed algorithms with that of the existing methods \cite{Cell-Free,chen2020structured}.
For our proposed TPA algorithms, we adopt $\kappa=2$ and $\Gc=75\%$. For the user group method of \cite{chen2020structured}, $T=16$ pilot dimension is chosen, and for the semi-random and the greedy algorithms in \cite{Cell-Free}, the pilot dimension is $T=15$ to best exploit the potential of their methods. It can be observed that our proposed sMWIM and greedy algorithms outperform all others in $90\%$-likely spectral efficiency, while the structured user group method has the best $10\%$-likely rate performance.

In Figure \ref{fig:50p_30p_CDF}, the sum rate performance versus the pilot dimension $T$ is considered for all pilot assignment algorithms. For our proposed algorithms, we also consider the different connectivity when $\Gc=30\%$, $50\%$, and $75\%$ with $\kappa=2$. For comparison, our proposed LRMC algorithm to find the pilot dimension is also considered to improve the semi-random scheme. We observe that the sMWIM algorithm with $\Gc=75\%$ has the highest sum rate when $T=20$, but when $T$ is small or large, it is outperformed by the structured policy \cite{chen2020structured}. The sMWIM algorithm with $\Gc=30\%$ performs well when $T$ is small, because the sparsity lends itself to a relatively more efficient pilot assignment given the limited number of training resource, but the performance is significantly degraded when $T$ becomes larger, due to the remaining interference that is not captured by $\Gc$. 
Remarkably, when $T$ is extremely small, the semi-random algorithm turns out to be the best choice. 
The structured policy with user group scheme has the superior sum rate performance if budget of pilot dimension is larger than 24, which is more than needed for our methods.
In addition, for our proposed sMWIM algorithm, when $T$ is small, then a sparser connectivity $\Gc$ yields a better sum rate performance; when $T$ exceeds certain threshold (e.g., $T=12$), then the denser the connectivity $\Gc$ is, the better the sum rate is. It suggests that if training resource is limited, a sparser $\Gc$ is preferable, and vice versa.
Our proposed greedy algorithm could also have a better sum rate performance if the pilot dimension is properly chosen, i.e., $T=20$.
As a side remark, our proposed methods do not require the prior knowledge of pilot dimension as the user group scheme does \cite{chen2020structured}.
The pilot dimension corresponding to the peak sum rate value indicates the minimum number of training dimensions for pilot assignment. We can observe that the training dimension of sMWIM increases with the density of network connectivity $\Gc$ -- it requires $T=20$, $T=16$, and $T=12$ for $\Gc=75\%$, $\Gc=50\%$, and $\Gc=30\%$, respectively. }
\begin{figure}
 \begin{minipage}{0.48\linewidth}
\centering
\includegraphics[width=1\columnwidth]{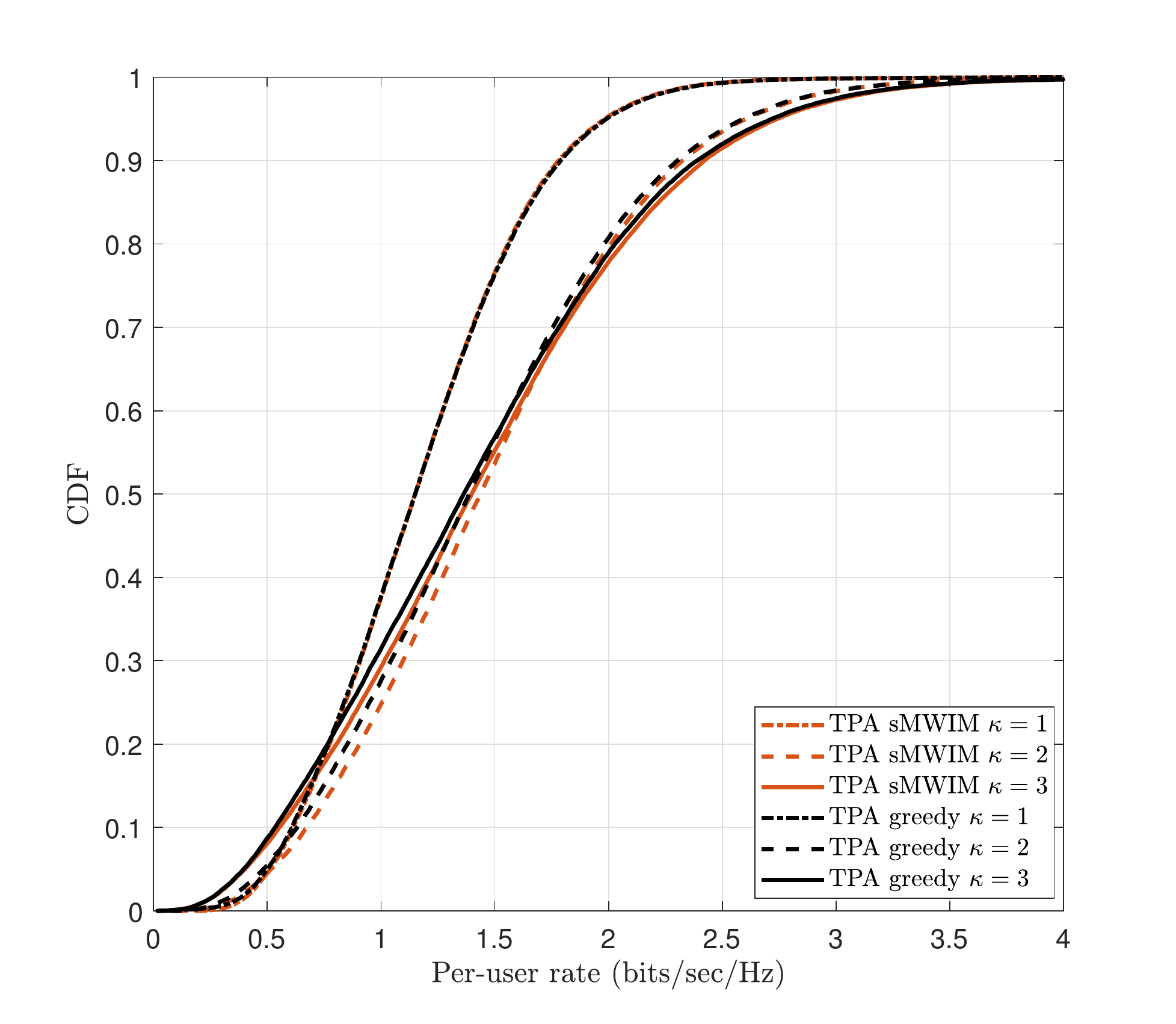}
\caption{The CDF of the downlink achievable rate per user with $\Gc=75\%$.}
\label{fig:30p_30p_CDF}
\end{minipage}
\quad
 \begin{minipage}{0.48\linewidth}
 \centering
\includegraphics[width=1\columnwidth]{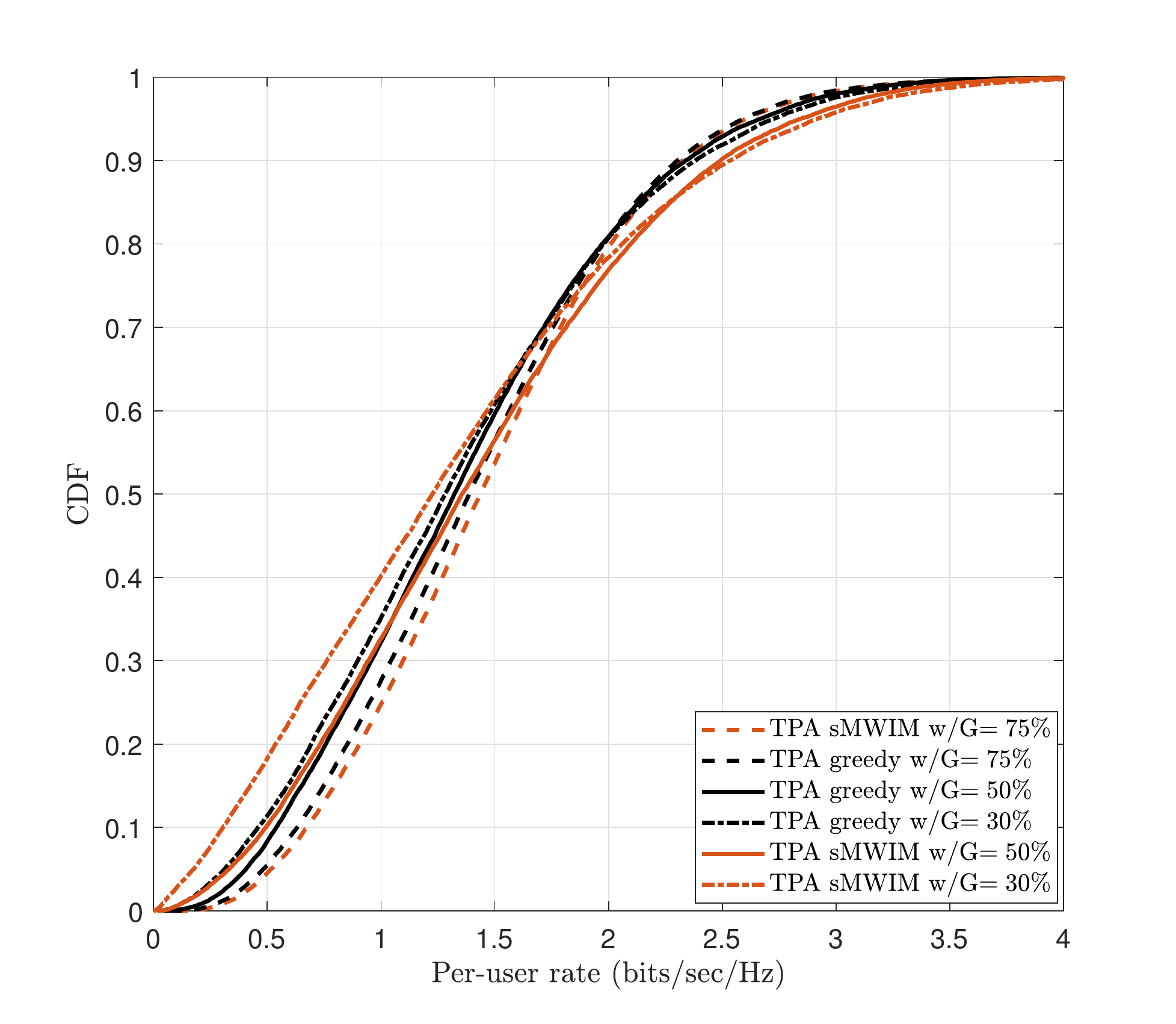}
\caption{The CDF of the downlink achievable rate per user with $\kappa=2$.}
\label{fig:30p_sumrate}
 \end{minipage}
 \vspace{-20pt}
\end{figure}

To evaluate the impact of $\kappa$ and $\Gc$, we plot the CDF of the downlink achievable rate with different $\kappa$ in Figure \ref{fig:30p_30p_CDF} and with different $\Gc$ in Figure \ref{fig:30p_sumrate}. Figure \ref{fig:30p_30p_CDF} illustrates the CDF of per-user rate performance of both sMWIM and greedy algorithm with $\kappa=1,2,3$ when $\Gc=75\%$ is fixed. We can observe that when $\kappa=1$, both sMWIM and greedy algorithms have the same performance. Note that $\kappa=1$ means each RRH is allowed to estimate the channel from one UE in each pilot dimension, so that the pilot dimension is minimized. As the pilot scheduling is on the artificially imposed structure $\Gc$, pilot contamination is inevitable and may not be necessarily eliminated perfectly in the physical scenarios. As such, by setting $\kappa=2,3$, certain level of pilot contamination is allowed in $\Gc$. In doing so, the majority of UEs witness certain increase in per-user rate performance, although there is some degradation of the UE with low rate. To summarize, $\kappa=2$ is preferred with respect to per-user rate performance, where a limited level of pilot contamination is allowed in pilot assignment.  
Figure \ref{fig:30p_sumrate} illustrates the CDF of the downlink per-user rate when different connectivity $\Gc$ is considered under a fixed $\kappa=2$. It can be observed that, for the sMWIM algorithm, when the connectivity is denser (e.g., $\Gc=75\%$), the 90\%-likely per-user rate is higher, as potential pilot contamination and multiuser interference is taken into account although there might be less freedom for pilot assignment. On the other hand, when the connectivity is sparser (e.g., $\Gc=30\%$) the 10\%-likely per-user rate is higher, meaning that there would be more UEs have per-user rate above 2.5 bits/sec/Hz. There observations agree on the intuition that a proper UE-RRH association is crucial for the sMWIM algorithm. For the greedy algorithm, the per-user rate performance is less sensitive to the connectivity $\Gc$. It is because in the greedy algorithm the network connectivity $\Gc$ will be refined before pilot assignment (see $\tilde{\Tm}^{(t)}$ in Alg. \ref{greedy}). We observe that the performance is slightly outperformed by the sMWIM algorithm. One reason is that, each UE is assigned with one unique orthogonal pilot in the greedy algorithm, while in the sMWIM algorithm the pilot of one UE could be the linear combination of multiple orthogonal pilots - this suggests the potential benefit of coded pilot design. Nevertheless, the computational complexity of the greedy algorithm is substantially reduced.

\section{Conclusion}
\label{sec:con}
We have proposed a framework for pilot assignment in large-scale distributed MIMO networks by artificially imposing topological structures on UE-RRH connectivity. By such a topological modeling, we cast the pilot assignment problem to a topological interference management (TIM) problem with groupcast messages. With respect to the known or unknown channel estimation patterns, we proposed two topological pilot assignment (TPA) problem formulations by a low-rank matrix completion and factorization method  and a binary quadratically constrained quadratic program, for which we apply low-complexity algorithms to solve the pilot assignment problem efficiently. The effectiveness of our proposed frameworks and algorithms are verified under the cell-free massive MIMO settings. The proposed TPA approach yields superior ergodic rate performance compared to the state-of-the-art pilot assignment methods.
The bridge between TPA and TIM problems is expected to trigger a new line of research dedicated to channel estimation methods in distributed networks. The rich coding tools from TIM will be hopefully tailored for pilot assignment applications in distributed MIMO systems.


\end{document}